# Border mapping multi-resolution (BMMR) technique for incompressible projection-based particle methods

Cezar Augusto Bellezi[a,b], Liang-Yee Cheng[b], Rubens Augusto Amaro Jr.[b], Marcio Michiharu Tsukamoto[b]

[a]Department of Naval and Ocean Engineering, Polytechnic School at the University of São Paulo, Av. Prof. Mello Moraes, 2231, Cidade Universitária, São Paulo, CEP 05508-030, São Paulo, SP, Brazil

[b]Department of Construction Engineering, Polytechnic School at the University of São Paulo, Av. Prof. Almeida Prado, trav. 2, 83 - Cidade Universitária, SP, 05508900, São Paulo, SP, Brazil

cezar.bellezi@usp.br, cheng.yee@usp.br, rubens.amaro@usp.br, michiharu@usp.br

**Abstract**

A novel multi-resolution technique called border mapping multi-resolution (BMMR) is proposed for projection-based particle methods. The BMMR aims to obtain background equivalent particle distributions in the two sides of a border between sub-domains with a 2:1 resolution ratio so that a single resolution framework is adopted to near-border particles calculations. The novelty of the BMMR is that it obtains the background grid from the mapping of the actual particle distribution in the border and, as a result, the location of the particles in the background grid exactly matches the location of the actual particles. In this way, such technique aims to reduce the error from the interpolation of the physical quantities in the background grid and to avoid sudden changes in the particle distribution that may lead to unstable local pressure calculation. In the coarse side of the border, the refinement is made by a triangulation and by placing fictitious particles in the midpoints of the triangles. In the fine side of the border, the derefinement is made by defining a set of fine particles that result in a particle distribution that best resembles a coarse one. The accuracy and computational performance of the BMMR implemented in a moving particle semi-implicit (MPS) simulation system are verified by using benchmark test cases of 2D free surface flows.

# 1 Introduction

The Lagrangian particle-based methods have been receiving ever increasing attention due to its capacity to handle very complex flow phenomena relevant for different fields of engineering, e.g., Gotoh & Khayyer (2018), Ye et al. (2019), Li et al. (2020), Lind et al. (2020) and Luo et al. (2021). Particle methods have the advantages of inherent mass conservation, free surface tracking without additional treatment, and easy modeling of fluid-structure interaction (FSI) problems involving large deformation, fragmentation and merging or multiphase flows (Lind, et al., 2020). Besides, the adoption of particle-based methods avoids some of the major drawbacks associated with the traditional Eulerian mesh-based methods, such as the considerable effort regarding mesh generation, the numerical diffusion of the advective term of the Navier-Stokes equation and the deterioration in the accuracy of the results as a consequence of large mesh deformations (Koshizuka, et al., 2018).

Two of the most widely adopted particle methods for fluid dynamics are smoothed particle hydrodynamics (SPH) and moving particle semi-implicit (MPS). The SPH originated from the context of astrophysics (Gingold & Monaghan, 1977; Lucy, 1977) and later was adapted for fluid dynamics (Monaghan, 1994). On the other hand, the MPS was initially proposed for the simulation of incompressible flows (Koshizuka & Oka, 1996). Despite the similarities between the SPH and MPS (Souto-Iglesias, et al., 2013), the spatial discretization formulation differs between them, as the SPH is based on the integral representation of physical quantities while the MPS is based on Taylor series expansions to approximate the differential operators in irregular particles distributions. In common, the original formulations of the SPH and the MPS and their up-to-date versions adopt a single uniform resolution in the entire computational domain. This feature aggravates the problem of the high computational cost of the particle-based methods, mainly when the regions of the flow with large gradients of velocity and pressure occur only in small and restricted regions but the entire domain should be modeled using the very fine resolution. This makes computationally unfeasible the simulation of multi-scale problems.

Generally, two different strategies are adopted to tackle such shortcoming. The first approach is the coupling between the particle method, which is used to simulate only the concerning region (near field), and other less computer-intensive method, which is adopted to model the remaining of the domain (far field), e.g., particle methods coupled to boundary element method (BEM) (Sun, et al., 2017), non-hydrostatic wave-flow model (Altomare, et al., 2018), finite volume method (FVM) (Chiron, et al., 2018; Di Mascio, et al., 2021), and non-linear potential flow (Verbrugghe, et al., 2018). The main challenge of such approach is to guarantee consistent coupling between the domains modeled by different methods when applied to violent free-surface motion. On the other hand, a second and more flexible approach is the multi-resolution particle modeling, in which the computational domain could be fulfilled by particles of different resolutions.

In the context of weakly-compressible particle methods, in which an equation of state and a fully explicit algorithm are adopted, several works regarding the weakly-compressible SPH (WCSPH) have been reported. Feldman & Bonet (2007) developed a two-dimensional (2D) multi-resolution algorithm with pre-defined *refinement zones*[1]. In the refinement algorithm, a *mother*[2] particle is split into a set of *daughter* particles with a non-uniform mass distribution calculated through a constrained minimization of the local density change. Triangular and hexagonal patterns are proposed for the refinement, in which four and seven *daughter* particles are generated, respectively. Based on the seminal ideas of Feldman & Bonet (2007) for the hexagonal splitting pattern and local density optimization, Vacondio et al. (2013) proposed a 2D multi-resolution WCSPH with pre-defined refinement zones and both splitting and coalescing schemes, which are suitable for cases involving flows in both directions of the subdomains. In the coalescing algorithm, two particles are merged into a larger one. Although well predicted results of velocity and pressure were computed by using a particle shifting correction, the size of the particles within a subdomain is expected to gradually change during the simulation, since the ratios of splitting and coalescing are different. This approach was later extended for a three-dimensional (3D) WCSPH formulation by Vacondio et al. (2016).

Reyes López et al. (2013) proposed a refinement scheme that generates a square pattern of four

---

[1] Spatial region where the resolution increase will take place.
[2] A large particle that is split into several smaller (*daughter*) particles following a predefined pattern.

*daughter* particles. The kinematics of the test cases presented good results, but no considerations about the pressure were provided. Furthermore, since the derefinement is not modeled, this approach is only appropriate for cases with particles flowing from the coarse to the fine domain. In Barcarolo et al. (2014), a 2D multi-resolution WCSPH with refinement and derefinement by means of the concept of adaptative particle refinement (APR) is proposed. The refinement is based on the square pattern from Reyes López et al. (2013), although the *mother* particle is kept after the splitting and follows the motion of the *daughter* particles passively. In the derefinement scheme, the *mother* particle is *re-activated* while the *daughter* particles are removed. The main advantage of this approach is that the ratios of refinement and derefinement are the same, so that the resolution of the subdomains is kept uniform. However, instantaneous pressure discontinuities in the border between the subdomains were reported ("refinement shock") and a "buffer zone" was introduced to provide a smoother transition. Besides, further investigation of the performance of the derefinement scheme in case of dispersion within the refinement zone of the *daughter* particles generated from a given *mother* particle may be necessary. Further, the APR concept was adopted by Chiron et al. (2018) for a 2D multi-resolution WCSPH. To mitigates the formation of well-defined structures in the flow, such as the particle clustering along streamlines, they included a particle disordering step to the *daughter* particles created by the refinement process proposed by Reyes López et al. (2013). Besides, they replaced the "buffer zone" adopted by Barcarolo et al. (2014) by *fictitious* interpolated particles (named "guard particles") in the border between the subdomains. The test cases presented a smooth pressure field and good agreement with experimental data. In Lyu et al. (2021), the improved APR technique by Chiron et al. (2018) was combined with a multi-phase SPH. Then, the model was applied to water entry problems and a systematic numerical analysis was conducted to investigate the cavity dynamics of a wedge penetrating fluid interfaces with different density-ratios. The square pattern refinement proposed by Reyes López et al. (2013) was adopted in Hu et al. (2020), and a 2D multi-resolution WCSPH scheme was proposed using the concept of hybrid particle interaction (HPI), a numerical scheme that provides a smooth transition zone, which comprises simultaneously *mother* and *daughter* particles. The particles of the coarse subdomain interact with the *mother* particles while the particles of the fine domain interact with the *daughter* particles. The numerical simulations of the study show a continuous pressure field between subdomains, but spurious pressure waves are present in the results of a water entry case. Recently, Yang et al. (2021) proposed a 2D multi-resolution WCSPH scheme based on the assumption that the fine resolution subdomain should be close to the free surface. The technique was called adaptative spatial resolution (ASR), in which coarse particles automatically split in two as they approach the free surface and fine particles automatically merge to another particle as they move away from the free surface, i.e., the diameters of the particles increase with the distance from the free surface. Moreover, particles with different sizes are merged within the fluid.

Regarding the incompressible projection-based particle methods that solve a pressure Poisson equation (PPE), most of the authors used the MPS, with few exceptions that applied the incompressible SPH (ISPH) (Cummins & Rudman, 1999; Lo & Shao, 2002). Tanaka et al. (2009) were the first to propose a 2D multi-resolution MPS scheme, by two different approaches. In the first one, refinement zones were pre-defined, and the resolution of the subdomains was maintained by particle splitting and coalescing. The kinematics of single and multi-resolution simulations agreed well, but spurious pressure oscillations were clearly observed from the multi-resolution simulations. In order to mitigates both the non-physical pressure oscillations and uneven particle distributions, Tanaka et al. (2018), and Liu & Zhang (2021) in the context of multiphase, applied least squares MPS (LSMPS) (Tamai & Koshizuka, 2014) to their multi-resolution scheme. The results presented remarkable improvements for the pressure field. However, it should be mentioned that the LSMPS is substantially more computer-costly as a 3x3 matrix should be inverted for each particle, at each step, in 2D simulations. In the second approach proposed by Tanaka et al. (2009), the resolution of the particles is defined only at the initial step and particles with different sizes freely mix as the simulation progresses. The same approach was adopted by Tang et al. (2016a) and Tang et al. (2016b) for 2D and 3D MPS, respectively, and Shobeiry & Ardakani (2019) for 2D ISPH. It should be mentioned that this approach has a limited scope of application, since it does not ensure that fine resolution particles remain in the targeted zone.

Afterwards, Shibata et al. (2012) developed a 2D multi-resolution MPS scheme, the so-called overlapping particle technique (OPT). In this approach, the entire computational domain is simulated using the coarse resolution, while only the targeted area is modeled by overlapped fine resolution particles. The boundaries of the fine resolution case are modeled as inlet-outlet boundaries in which velocity and pressure fields are interpolated from the coarse resolution simulation. As advantages, such approach allows the use of the simpler single-resolution formulation of the discrete differential operators, and it also easily maintains the resolution of the subdomains uniform. On the other hand, some disadvantages consist of the weak-form, one-way coupling and additional treatment to compensate the mass loss at the inlet-outlet boundaries. Later, Tang et al. (2016c) extended the OPT for 3D cases and Shibata et al. (2017) developed a two-way formulation.

A 2D multi-resolution MPS scheme based on the no surface detection (NSD) (Chen, et al., 2014) technique with pre-defined refinement zones was further introduced by Chen et al. (2016). When a particle crosses the subdomains, particle splitting or merging is applied. In the splitting scheme, the *mother* particle is split into seven *daughter* particles in a hexagonal arrangement. The merging scheme joins two particles into one, so that the resolution within the subdomains gradually changes as simulation progresses due to the mismatch between the splitting ratio and the merging ratio.

Tsuruta et al. (2016) established a 2D multi-resolution projection-based particle method with a novel weight function for improved calculation of the interparticle forces and a novel particle number density model based on the concept of the potential density. Besides, they used the so-called space potential particle (SPP) (Tsuruta, et al., 2015), in which virtual particles are introduced to compensate the voids that may occur due to irregular arrangement of particles with different sizes. The fine resolution of the targeted area is obtained by particle splitting and no particle merging algorithm is adopted. Further, Khayyer et al. (2019) used this framework to model FSI problems in which the fluid is modeled by using a coarse resolution and the elastic structures are modeled by using fine resolution particles.

Although several efforts have been undertaken towards the development of the multi-resolution techniques, some key challenges remain, specially for incompressible projection-based particle methods. Compared to weakly-compressible methods, incompressible projection methods generally yield better in terms of pressure calculation and volume conservation (Lee, et al., 2008; Gotoh, et al., 2013). However, in the projection particle-based methods, the source term of the PPE usually contains a density invariant term so that the development of multi-resolution techniques is a much more complex task. Especially, the local particle density variations caused by splitting, coalescing, refinement or derefinement schemes may lead to poor numerical stability, with spurious local pressure oscillations. In order to elucidate these challenges, it is useful to make an analogy between multi-resolution techniques and the coupling between subdomains with different resolutions. While in the weakly compressible methods the conservation equations of each subdomain can be solved separately and sequentially, in the incompressible projection-based methods, ideally, such *coupling* should be in the strong-form, i.e., the conservation equations should be solved simultaneously with the information propagating instantaneously in both domains. Besides, as the particles flow from one subdomain to other with different resolution, the algorithm should update the resolution of the particles smoothly, to avoid abrupt changes in particle density. As an additional desirable feature, the resolution within a subdomain should be maintained as uniform as possible.

In light of the above, the contribution of the present work is a novel multi-resolution technique for projection-based particle methods, herein denominated as **B**order **M**apping **M**ulti-**R**esolution (BMMR). The BMMR technique divides the computational domain in subdomains with different resolutions and aims to obtain a background equivalent particle distribution in each side of the border between the subdomains so that the computation of velocity and pressure of a particle in the vicinity of the border can consider exclusively a single resolution particle distribution. The algorithm of the BMMR technique is characterized by two tasks: the simplification (derefinement) of the near border fine resolution particle distribution and the refinement of the near border coarse resolution particle distribution. With that, the formulation for strong coupling between the subdomains with different particle sizes can be easily derived while the resolution within each subdomain can be maintained uniform. Moreover, the technique is applied to the projection-based formulation of the MPS, in which a semi-implicit algorithm solves the governing equation, herein called as border mapping multi-resolution MPS (BMMR-MPS). In this way, the strong coupling of uniform-resolution subdomains for the projection-based MPS make the BMMR-MPS distinct from the models mentioned previously. Table 1 summarizes the main characteristics of the previous multi-resolution techniques and the proposed BMMR-MPS. Compared with other projection-based multi-resolution methods, the set of features covering strong coupling between uniform-resolution subdomains, which might be confined respectively in desirable positions avoiding disordered mix of particles with different resolutions, global conservation of mass and energy, and numerical accuracy, are the major advantages of the BMMR.

Furthermore, instead of split-merge procedures or background grid techniques that replace actual irregular particle distributions by background regular lattices, which are commonly adopted in the previous works, the BMMR obtains a background grid from the mapping of the actual particle distribution in the border and, as a result, the location of the particles in the background grid exactly matches the location of the actual particles. In this way, the BMMR generates a more consistent background grid in the case of irregular particle distributions, which avoids the abrupt local oscillations of the particle density, as indirectly supported by the density error results from Vacondio et al. (2016), and consequently unstable local pressure calculation is mitigated.

The BMMR-MPS is analyzed considering different free surface benchmark test cases, with the aim of highlighting the strengths and weaknesses of the approach concerning the accurate prediction of the flow evolution and pressure field, as well as conservation of the global properties such as mass and mechanical energy. Moreover, computational time to solve the PPE is compared between the BMMR-MPS and fine single-resolution MPS.

Table 1 - Main characteristics of previous multi-resolution techniques and the proposed BMMR-MPS.

| | Particle method | Pressure calculation | Coupling | Refinement (volume ratio) | Derefinement (volume ratio) | Subdomain resolution | Support radius | Additional treatments |
|---|---|---|---|---|---|---|---|---|
| (Feldman & Bonet, 2007) | SPH 2D | Explicit | Strong | Splitting 1:4 and 1:7 | None | Uniform | Proportional to resolution | - |
| (Tanaka, et al., 2009) | MPS 2D | Implicit | Strong | Splitting 1:2 and 1:4 | Merging 2:1 | Variable | Mean of pair of particles | Modified operators considering particle sizes |
| (Shibata, et al., 2012) | MPS 2D | Implicit | Weak one-way | Inlet/Outlet | Inlet/Outlet | Uniform | Proportional to resolution | Corrected inlet-outlet position/velocity |
| (Reyes López, et al., 2013) | SPH 2D | Explicit | Strong | Splitting 1:4 | None | Uniform | Proportional to resolution | - |
| (Vacondio, et al., 2013) | SPH 2D | Explicit | Strong | Splitting 1:7 | Merging 2:1 | Variable | Proportional to resolution | Artificial diffusion, Particle shifting |
| (Barcarolo, et al., 2014) | SPH 2D | Explicit | Strong | Splitting 1:4 | Passive following | Variable | Proportional to resolution | Artificial diffusion, Transition zone |
| (Vacondio, et al., 2016) | SPH 3D | Explicit | Strong | Splitting 1:8 to 1:21 | Merging 2:1 | Variable | Proportional to resolution | Artificial diffusion, Particle shifting |
| (Chen, et al., 2016) | MPS 2D | Implicit | Strong | Splitting 1:7 | Merging 2:1 | Variable | Unique for all domain | No Surface Detection (NSD), multi-step splitting |
| (Tang, et al., 2016a) | MPS 2D | Implicit | Weak one-way | Inlet/Outlet | Inlet/Outlet | Uniform | Proportional to resolution | Corrected inlet-outlet position/velocity |
| (Tang, et al., 2016b) | MPS 3D | Implicit | Strong | None | None | Variable | Mean of pair of particles | Modified operators considering particle sizes |
| (Tang, et al., 2016c) | MPS 3D | Implicit | Strong | None | None | Variable | Mean of pair of particles | Modified operators considering particle sizes |
| (Tsuruta, et al., 2016) | MPS 2D | Implicit | Strong | Splitting 1:2, 1:3, 1:4 | None | Variable | Unique for all domain | Space Potential Particle (SPS), Modified weight and PND |
| (Shibata, et al., 2017) | MPS 2D/3D | Implicit | Weak two-way | Inlet/Outlet | Inlet/Outlet | Uniform | Proportional to resolution | Corrected inlet-outlet position/velocity |
| (Tanaka, et al., 2018) | MPS 2D | Implicit | Strong | Splitting 1:4 | Passive following | Variable | Mean of pair of particles | Higher order operators, Packing ratio for incompressible constraint |
| (Chiron, et al., 2018) | SPH 2D | Explicit | Strong | Splitting 1:2 and 1:4 | Merging 2:1 | Variable | Proportional to resolution | Prolongation and Restriction |
| (Khayyer, et al., 2019) | MPS 2D | Implicit | Strong | None | None | Uniform | Unique for all domain | Modified weight and PND |
| (Hu, et al., 2020) | SPH 2D | Explicit | Strong | Splitting 1:4 | None | Uniform | Proportional to resolution | Artificial diffusion, Transition zone |
| (Yang, et al., 2021) | SPH 2D | Explicit | Strong | Splitting 1:2 | Merging 2:1 | Variable | Mean of pair of particles | Artificial diffusion |
| BMMR-MPS | MPS 2D | Implicit | Strong | Triangulation 1:4 | Simplification 4:1 | Uniform | Proportional to resolution | Transition zone for the source term of PPE |

## 2   Moving particle semi-implicit (MPS) method

The governing equations for incompressible viscous flow can be written in a moving Lagrangian frame as follows:

$$\frac{D\rho}{Dt} = -\rho \nabla \cdot \mathbf{u} = 0, \tag{1}$$

$$\frac{D\mathbf{u}}{Dt} = -\frac{\nabla P}{\rho} + \nu \nabla^2 \mathbf{u} + \mathbf{f}_{\text{ext}} + \mathbf{g}, \tag{2}$$

where $\rho$ is the fluid density, $\mathbf{u}$ denotes the velocity vector, $P$ represents the pressure, $\nu$ stands for the kinematic viscosity, $\mathbf{f}_{\text{ext}}$ is the vector of the external body force per unit mass and $\mathbf{g}$ is the gravitational field.

In the MPS method, the spatial differential operators are replaced by discrete operators of a given particle $i$, $\langle \ \rangle_i$, derived from a weight function $W_{ij}$ that accounts for the influence of a neighbor particle $j \in \Omega_i$. Here we adopted a non-singular second-order polynomial weight function (Kondo & Koshizuka, 2011), given by:

$$W_{ij} = \begin{cases} \left(\frac{\|\mathbf{r}_{ij}\|}{r_e} - 1\right)^2 & \|\mathbf{r}_{ij}\| \leq r_e \\ 0 & \text{otherwise} \end{cases}, \tag{3}$$

where $r_e$ is the effective radius that limits the range (compact support size) of the neighborhood $\Omega_i$ of the particle $i$ and $\|\mathbf{r}_{ij}\| = \|\mathbf{r}_j - \mathbf{r}_i\|$ represents the distance between the particles $i$ and $j$. Following the recommendations from Koshizuka and Oka (1996), in the present work the *small* effective radius $r_e = 2.1 \times l_i$ is used for the first-order differential operators, such as gradient operator (see Eq. (4)) and divergence operator (see Eq. (5)), and also for the particle number density (see Eq. (7)). On the other hand, the *large* effective radius $r_e = 3.1 \times l_i$ is used for the second-order differential operators, e.g., the Laplacian operator (see Eq. (6)). $l_i$ is the initial distance between particles.

It should be highlighted that, in this work, the entire domain of actual particles $\mathbb{P}$ is classified as inner fluid $\mathbb{I}$, free surface $\mathbb{F}$, wall $\mathbb{W}$ and dummy $\mathbb{D}$, i.e., $\mathbb{P} = \mathbb{I} \cup \mathbb{F} \cup \mathbb{W} \cup \mathbb{D}$.

For an arbitrary scalar function $\phi$ and an arbitrary vector $\boldsymbol{\phi}$, the spatial differential operators,

e.g., gradient, divergence and Laplacian, are approximated by:

$$\langle \nabla \phi \rangle_i = \frac{dim}{n^0} \sum_{j \in \Omega_i} \frac{\phi_j - \phi_i}{\|\mathbf{r}_{ij}\|^2} \mathbf{r}_{ij} W_{ij} , \qquad (4)$$

$$\langle \nabla \cdot \boldsymbol{\phi} \rangle_i = \frac{dim}{n^0} \sum_{j \in \Omega_i} \frac{\boldsymbol{\phi}_j - \boldsymbol{\phi}_i}{\|\mathbf{r}_{ij}\|^2} \cdot \mathbf{r}_{ij} W_{ij} , \qquad (5)$$

$$\langle \nabla^2 \phi \rangle_i = \frac{2 dim}{\lambda^0 n^0} \sum_{j \in \Omega_i} (\phi_j - \phi_i) W_{ij} , \qquad (6)$$

where $dim$ is the number of spatial dimensions and $n^0$ denotes the constant particle number density for a fully filled compact support. The particle number density $n_i$, which is proportional to the fluid density $\rho$, is defined by:

$$n_i = \sum_{j \in \Omega_i} W_{ij} . \qquad (7)$$

The constant $\lambda^0$, is a correction parameter so that the variance increase is equal to that of the analytical solution, and is computed considering a fully filled compact support by:

$$\lambda^0 = \frac{\sum_{j \in \Omega_i} W_{ij}^0 \|\mathbf{r}_{ij}^0\|^2}{\sum_{j \in \Omega_i} W_{ij}^0} . \qquad (8)$$

## 2.1 Particle regularization techniques

### 2.1.1 Particle shifting (PS)

To improve the uniformity of the particle distribution, the particle shifting (PS) technique governed by Fick's law of diffusion, as proposed by Lind et al. (2012) in the ISPH context, is adopted here:

$$\Delta \mathbf{r}_i = \begin{cases} -A_F (l_i)^2 C_r M_a \langle \nabla C \rangle_i & i \in \mathbb{I} \\ 0 & i \in \mathbb{F} \end{cases} , \qquad (9)$$

where $i \in \mathbb{I}$ represents the inner fluid particles, $i \in \mathbb{F}$ is the free surface particles, $A_F \in [1, 6]$ and $M_a$ denotes the Mach number. We adopted $A_F = 2$ for all simulations, as recommended in Skillen et al. (2013).

To avoid extreme deviation of the particle position, the magnitude of $\Delta \mathbf{r}_i$ is limited here by:

$$\Delta \mathbf{r}_i = \min(0.05 \times l_i, \|\Delta \mathbf{r}_i\|) \frac{\Delta \mathbf{r}_i}{\|\Delta \mathbf{r}_i\|}. \tag{10}$$

The gradient of the concentration $C_i$ (volume fraction) is obtained by (Jandaghian & Shakibaeinia, 2020):

$$\langle \nabla C \rangle_i = \frac{dim}{n^0} \sum_{j \in \Omega_i} \frac{C_i + C_j}{\|\mathbf{r}_{ij}\|^2} \mathbf{r}_{ij} \omega_{ij}, \tag{11}$$

with

$$C_i = \frac{\sum_{j \in \Omega_i} W_{ij}}{n^0}. \tag{12}$$

### 2.1.2 Particle collision (PC)

The pressure gradient is not properly computed for particles with truncated support, i.e., free-surface particles $i \in \mathbb{F}$. In this way, a particle collision (PC) model is highly recommended to adjust the distances between particles, then avoiding numerical instabilities related to particle clustering. Here, a pair-wise PC model is applied after the explicit first stage of the MPS method, and a repulsive velocity vector for overlapped inner fluid particles is enforced by:

$$\Delta \mathbf{u}_i = \begin{cases} \sum_{j \in \Omega_i} \frac{(1+\alpha_2)}{\alpha_3} \frac{\mathbf{r}_{ij} \cdot \mathbf{u}_{ij}}{\|\mathbf{r}_{ij}\|} \frac{\mathbf{r}_{ij}}{\|\mathbf{r}_{ij}\|} & \|\mathbf{r}_{ij}\| \leq \alpha_1 l_i \text{ and } \mathbf{r}_{ij} \cdot \mathbf{u}_{ij} < 0 \\ 0 & \text{otherwise} \end{cases}, \tag{13}$$

where $\mathbf{u}_{ij} = \mathbf{u}_j - \mathbf{u}_i$, and values of $\alpha_1 \in [0.8, 1.0]$ and $\alpha_2 \in [0.0, 0.2]$ enhance the spatial stability (Lee, et al., 2011). For all cases analyzed herein, we adopt $\alpha_1 = 0.85$ and $\alpha_2 = 0.2$. If the neighbor particle is an inner fluid or free surface particle ($j \in \mathbb{I} \cup \mathbb{F}$) then $\alpha_3 = 2$, otherwise ($j \in \mathbb{W} \cup \mathbb{D}$) $\alpha_3 = 1$.

## 2.2 Boundary conditions

### 2.2.1 Free surface

The neighborhood particles centroid deviation (NCPD) technique (Tsukamoto, et al., 2016) is adopted to identify the free-surface particles $i \in \mathbb{F}$, and its pressure is set to zero $P_{i \in \mathbb{F}} = 0$ following the Dirichlet dynamic boundary. The first step of NPCD has a low computational cost and gives a rough detection of the free-surface particles, as follows:

$$\begin{cases} n_i < \beta_F \cdot n^0 & \to \quad i \in \mathbb{F} \\ \text{otherwise} & \to \quad i \in \mathbb{I} \end{cases}, \tag{14}$$

where $i \in \mathbb{I}$ represents the inner fluid particles. After that, a more precise detection is performed only on free-surface particles $i \in \mathbb{F}$, and they are finally classified as free surface by:

$$\begin{cases} \sigma_i > \varrho_F \cdot l_i \text{ and } N_i \leq 4 & \to \quad i \in \mathbb{F} \\ \text{otherwise} & \to \quad i \in \mathbb{I} \end{cases}, \tag{15}$$

where $N_i$ represents the number of neighbors of particle $i$.

The deviation $\sigma_i$ is calculated as:

$$\sigma_i = \frac{\sqrt{\left[\sum_{j \in \Omega_i} W_{ij}(x_j - x_i)\right]^2 + \left[\sum_{j \in \Omega_i} W_{ij}(z_j - z_i)\right]^2 + \left[\sum_{j \in \Omega_i} W_{ij}(z_j - z_i)\right]^2}}{\sum_{j \in \Omega_i} W_{ij}}. \tag{16}$$

Koshizuka & Oka (1996) and Tsukamoto et al. (2016) suggest respectively $\beta_F \in [0.8, 1.0[$ and $\varrho_F \in [0.2, \infty[$. For all simulations performed herein, we adopted $\beta_F = 0.93$ and $\varrho_F = 0.25$.

### 2.2.2 Rigid wall

The representation of rigid wall boundary condition is done by imposing layers of solid particles. The particles that form the layer in contact with the fluid are denominated wall particles $\mathbb{W}$, of which the pressure is computed by solving PPE linear system (see Eq. (25)), together with the inner fluid particles $i \in \mathbb{I}$. The remaining layers are composed by dummy particles $\mathbb{D}$, which are used to assure the correct calculation of the particle number density of the wall particles.

The nonhomogeneous Neumann boundary condition of pressure is applied at rigid walls. Then,

the pressure of dummy particles $j \in \mathbb{D}$ can be approximated by:

$$P_{j \in \mathbb{D}} = P_i + \|\mathbf{r}_{ij}\| \frac{\partial P}{\partial n}\bigg|_{i \in \mathbb{W}}, \qquad (17)$$

with the following simplified relation (Matsunaga, et al., 2020):

$$\frac{\partial P}{\partial n}\bigg|_{i \in \mathbb{W}} = \rho \mathbf{n}|_{i \in \mathbb{W}} \cdot \mathbf{g} \approx \rho \frac{\mathbf{r}_{ij}}{\|\mathbf{r}_{ij}\|} \cdot \mathbf{g}. \qquad (18)$$

Eq. (17) is included in the PPE (see Eq. (25)) where the second term $\|\mathbf{r}_{ij}\| \frac{\partial P}{\partial n}\big|_{i \in \mathbb{W}}$ can be moved to the right-hand side, i.e., added to the source term of the linear system.

## 2.3 Algorithm

To solve the incompressible viscous flow, a semi-implicit algorithm is used in the MPS method, which is similar to the projection method (Harlow & Welch, 1965; Chorin, 1967; Temam, 1969). After the simplification (Section 3.1.1) and refinement (Section 3.1.2) of the border mapping technique, the particle shifting vector $\Delta \mathbf{r}_i^t$ is imposed to inner fluid particles $i \in \mathbb{I}$, and the position is adjusted as:

$$\mathbf{r}_i' = \mathbf{r}_i^t + \Delta \mathbf{r}_i^t. \qquad (19)$$

After that, predictions of velocity and position of the fluid particles ($i \in \mathbb{I} \cup \mathbb{F}$) are carried out explicitly by using viscosity and external forces terms of the momentum conservation (Eq. (2)):

$$\mathbf{u}_i^* = \mathbf{u}_i^t + [\nu \langle \nabla^2 \mathbf{u} \rangle_i + \mathbf{f}_i]^t \Delta t, \qquad (20)$$

$$\mathbf{r}_i^* = \mathbf{r}_i' + \mathbf{u}_i^* \Delta t. \qquad (21)$$

The Laplacian of the velocity in Eq. (20) is calculated as follows:

$$\langle \nabla^2 \mathbf{u} \rangle_i = \frac{2 dim}{\lambda^0 n^0} \sum_{j \in \Omega_i} \mathbf{u}_{ij} W_{ij}. \qquad (22)$$

Then the collision model is applied, following Eq. (13), and the contribution of $\Delta \mathbf{u}_i^*$ is added to the particle velocities and positions:

$$\mathbf{u}_i^{**} = \mathbf{u}_i^* + \Delta \mathbf{u}_i^*, \qquad (23)$$

$$\mathbf{r}_i^{**} = \mathbf{r}_i^* + \Delta \mathbf{u}_i^* \Delta t. \qquad (24)$$

Next, the pressures of all inner fluid and wall particles are calculated by solving the PPE, a linear system of algebraic equations, considering the source term linked to the particle number density (PND) criterion and the divergence of the velocity field, similar to that proposed by Tanaka e Matsunaga (2010), but using the so-called time-scale correction of particle-level impulses (TCPI) source term (Cheng, et al., 2021).

$$\langle \nabla^2 P \rangle_i^{t+\Delta t} - \frac{\rho}{\Delta t^2} \alpha_c P_i^{t+\Delta t} = c_s^2 \frac{\rho}{l_i^2} \left( \frac{n^0 - n_i^{**}}{n^0} \right) + c_s \frac{\rho}{l_i} \langle \nabla \cdot \mathbf{u} \rangle_i^{**}, \qquad (25)$$

where $n_i^{**}$ is the particle number density calculated based on the displacement of particles obtained in the explicit calculations, $c_s$ represents a relaxation parameter proportional to the propagation of speeds, and $\alpha_c$ is the coefficient of artificial compressibility.

The idea behind the adoption of relaxation parameter $c_s$ is to enforce the incompressibility condition in a robust way, while mitigating spurious oscillations in the discrete PPE. Henshaw & Kreiss (1995) and Li (2020) showed that the accuracy of a numerical method can be assured by adopting a properly tuned relaxation parameter, when a split-step strategy is used in the same manner as projection methods. The coefficient of artificial compressibility $\alpha_c$ makes the diagonal elements of the matrix bigger, rending it very useful for computational stabilization, i.e., improving the conditioning of the linear system of algebraic equations (Riley, 1955), here specifically the PPE. Nevertheless, both $c_s$ and $\alpha_c$ should be chosen appropriately in order to avoid non-physical fluid behavior. Typically, the values $c_s = A_{cs}\sqrt{gl_i}$, with $A_{cs} \in [1,30]$ (Cheng, et al., 2021) and $\alpha_c \in [10^{-9}, 10^{-8}]$ ms$^2$/kg (Arai, et al., 2013; Shibata, et al., 2015; Duan, et al., 2019; Tsukamoto, et al., 2020), provide stable simulations.

With respect to the divergence of the velocity in Eq. (25), it is approximated by:

$$\langle \nabla \cdot \mathbf{u} \rangle_i = \frac{dim}{n^0} \sum_{j \in \Omega_i} \frac{\mathbf{u}_{ij} \cdot \mathbf{r}_{ij}}{\|\mathbf{r}_{ij}\|^2} W_{ij} \,. \tag{26}$$

Finally, velocity and position are updated by a simple 1$^{st}$ order Euler integration:

$$\mathbf{u}_i^{t+\Delta t} = \mathbf{u}_i^{**} - \frac{\Delta t}{\rho} \langle \nabla P \rangle_i^{t+\Delta t} \,, \tag{27}$$

$$\mathbf{r}_i^{t+\Delta t} = \mathbf{r}_i^{**} + \left(\mathbf{u}_i^{t+\Delta t} - \mathbf{u}_i^{**}\right)\Delta t \,. \tag{28}$$

To prevent particle clustering, avoiding unstable behavior when attracting forces act between particles, the pressure gradient in Eq. (27) can be calculated as (Wang, et al., 2017):

$$\langle \nabla P \rangle_i = \left[\sum_{j \in \Omega_i} \frac{\mathbf{r}_{ij}}{|\mathbf{r}_{ij}|} \otimes \frac{\mathbf{r}_{ij}^T}{|\mathbf{r}_{ij}|} W_{ij}\right]^{-1} \sum_{j \in \Omega_i} \frac{P_j - \hat{P}_i}{\|\mathbf{r}_{ij}\|^2} \mathbf{r}_{ij} W_{ij} \,, \tag{29}$$

where $\hat{P}_i = \min_{j \in \Omega_i}(P_j, P_i)$. Notwithstanding, a relevant point is that the linear momentum is not conserved in Eq. (29), since the resulting interparticle pressure forces are not anti-symmetric (equal in magnitude, opposite in direction) (Khayyer & Gotoh, 2013).

A fixed time step $\Delta t$ is initially assigned following the CFL condition (Courant, et al., 1967):

$$\Delta t \leq \frac{C_r \, l_f}{|u|_{\max}} \,, \tag{30}$$

where $C_r \in \,]0, 1.0]$ denotes the Courant number, $l_f$ stands for the initial particle distance in the fine subdomain and $|u|_{\max}$ is the maximum flow velocity.

## 3 Border Mapping Multi-Resolution (BMMR) technique

In the context of multi-resolution techniques for projection-based particle methods, the desirable features are as follows:

1. absolute control of the extension of the fine and coarse resolution subdomains, confining it within the desirable regions of interest with relevant local details, and to be able to maintain the uniform resolution within the subdomains during the entire simulation;

2. the coupling between the subdomains with different resolutions should be in the strong-form, i.e., the conservation equations are solved jointly, with all the subdomains forming a monolithic system;

3. the algorithms should avoid abrupt changes of the local particle density in the border between the subdomains to provide more stable calculations;

4. the numerical model should be consistent with the Newton's third law to ensure the conservation of momentum independent of the particle sizes.

The mathematical models and numerical schemes of BMMR-MPS, which aims to provide such features, are presented and discussed in detail in the following sections.

## 3.1 Equivalent particle distribution

In the BMMR-MPS, the computational domain is divided into pre-defined refinement zones, or subdomains, of different resolutions. For sake of simplicity and proof of the concept, subdomains delimited by a polygon without self-intersection is considered herein despite the concept might also be applied to curved borders. In the interior of each subdomain, the neighborhood of a particle only contains particles of the same resolution so that the formulation of the original single-resolution MPS can be adopted directly. However, for the particles close to the border of the subdomains, their neighborhood may contain particles of an adjacent subdomain with a different resolution (Figure 1-a). In order to address such issue, the main concept of the BMMR-MPS is to obtain an "background equivalent particle distribution" in both sides of the border between the subdomains. A background equivalent fine resolution particle distribution is generated from the coarse particles in the near-border region by a refinement algorithm. Meanwhile, a background equivalent coarse resolution particle distribution is generated from the fine particles in the near border by a derefinement algorithm. In this way, the neighborhood of the near-border coarse resolution particles consider the background equivalent coarse particle distribution instead of the actual fine resolution particle distribution in the other subdomain (Figure 1-b) while the neighborhood of the near-border fine resolution particles consider the background equivalent fine particle distribution instead of the actual coarse resolution particle distribution in the other subdomain (Figure 1-c). By using the background equivalent particle distributions in the near-border regions, the single-resolution framework of the MPS could also be applied directly to the near border particles without any additional treatment or particle splitting or merging techniques, nor the discrete operators are required to consider particles of different resolutions.

The multi-resolution techniques for particle methods previously proposed in the literature adopt a strictly Lagrangian point-of-view, in which the particles correspond to macro-scale volumes of fluid that carry its own physical properties. By assuming that the Eulerian and Lagrangian points-of-view are equivalent at a given instant, it is useful to adopt a Eulerian point-of-view in order to better understand the refinement and derefinement algorithms of the BMMR-MPS. In this Eulerian point-of-view, the particle-based discrete domain could be depicted as an unstructured grid in which the nodes are placed at the center of the particles and they are the points in which the physical quantities are known. Such unstructured grid has an irregular arrangement of nodes that are almost uniformly spaced out. In such context, it is possible to adopt triangulation algorithms to create a triangular mesh from the particle distribution which, ideally, may comprise mainly equilateral triangles. Next, the triangles are divided by the refinement algorithm or merged by the derefinement algorithm in the near-border region. The nodes of the newly generated triangular meshes are then adopted to create the equivalent particle distributions in the near-border region.

Following this concept for the implementation of the proposed multi-resolution technique, the ratio of the distance between particles of two adjacent subdomains is always 2:1 (volume ratio 1:4 in 2D). Therefore, as the ratio of refinement and derefinement are the same, the BMMR-MPS is able to maintain constant and uniform particle resolution within a subdomain during the entire simulation. In order to achieve larger resolution ratios, successive subdomain refinements, which is a topic for future investigation, might also be considered.

Other key aspect of the refinement and the derefinement algorithms adopted by the BMMR-MPS technique is that the position of the particles in the background equivalent particle distributions matches the position of actual particles near the border. In the case of the derefinement algorithm, the position of the equivalent coarse particles must match the positions of some actual fine particles (Figure 1-b). In the case of the refinement algorithm, the position of some equivalent fine particles also matches the position of actual coarse particles (Figure 1-c). Without suddenly replacing a *mother* particle by *daughter* particles, such as in the commonly adopted splitting algorithms, the abrupt change in the particle distribution can be mitigated in the BMMR-MPS and a smoother distribution of particle density can be obtained near the border, which may contribute to more stable pressure calculations.

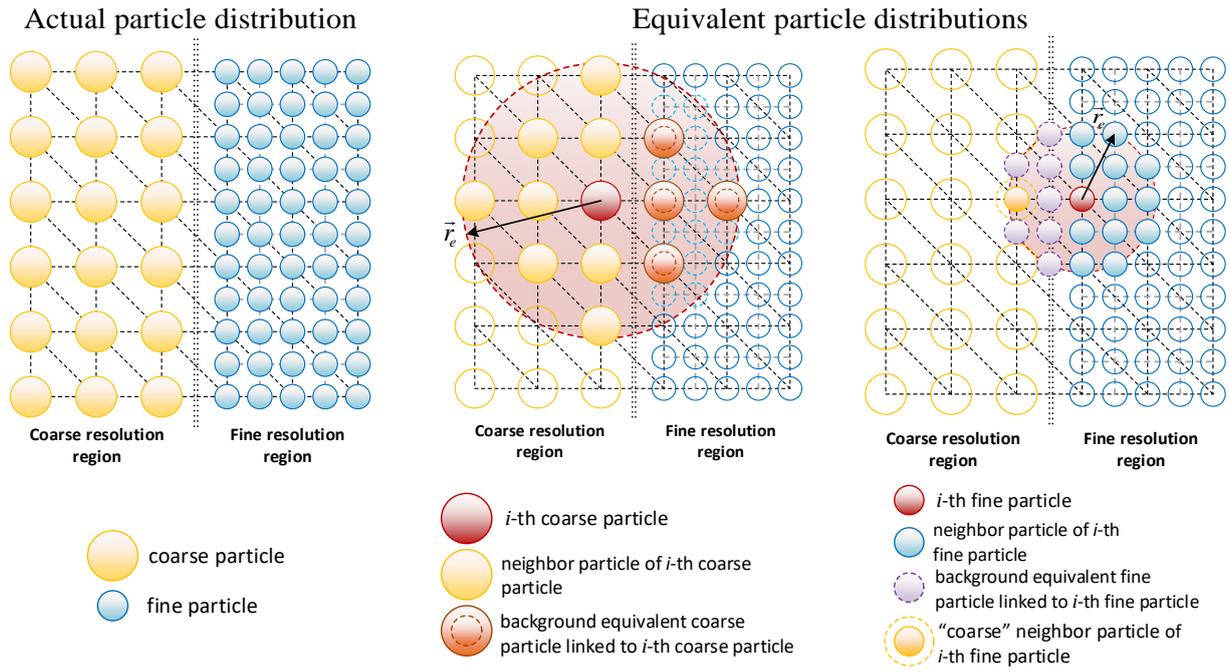

(a) Particles in the near-border region  (b) Background equivalent coarse particle  (c) Background equivalent fine particle

Figure 1 – Conceptual scheme of the border mapping multi-resolution (BMMR) technique – (a) actual particle distribution in the border, (b) neighborhood of *i*-th coarse resolution particle with its background equivalent coarse particles and (c) neighborhood of a *i*-th fine resolution particle with its background equivalent fine particles, both close to the border.

### 3.1.1 Derefinement: Simplification

The objective of the derefinement algorithm is to obtain a background equivalent coarse resolution particle distribution from the fine resolution particles in the near-border region (see Figure 2-a). Here, the derefinement algorithm is named as "simplification" based on the terminology adopted in computer graphics by algorithms that reduce the number of elements of a given mesh. The simplification scheme of the present work basically classifies the near-border fine resolution particles into two types: *simplified* and *non-simplified* particles. Then, the background equivalent coarse particle distribution is composed only by the fine resolution *simplified* particles (orange particles in Figure 1-b), of which the positions are considered as coarse particles ones in this context. The simplification of the near-border fine resolution particle distribution is carried out in two main parts. The first part consists in the simplification of the solid wall and dummy particles (Figure 2), which usually have a square lattice arrangement and, therefore, require a simpler approach. The second part is the simplification of the fluid particles (Figure 3), which generally have an irregular arrangement that requires a more complex derefinement algorithm.

The simplification algorithm starts with the identification of the set of fine particles that need to be classified into *simplified* and *non-simplified* ones, herein denominated as *near-border* particles. For this purpose, only the fine particles within a distance $r_{e,C} = 3.1 \times l_{i,C}$ of the border are considered (fine particles between the dashed lines in the right-hand side of Figure 2-a). In addition, the coarse particles within a distance of $l_{i,C}$ of the border are used as *reference* particles for the simplification process, where $l_{i,C}$ is the initial distance between coarse resolution particles (highlighted coarse particles between the dashed lines in the left-hand side of the border in Figure 2-a). After defining the *near-border* particles, an iterative simplification process is adopted. As the rigid walls are commonly modeled by a regular square particle arrangement, the wall particles are a suitable and easy spot to start the iterative simplification process. First, the fine resolution wall particles at a distance of $l_{i,F}$ to $\sqrt{2} \times l_{i,F}$ to the rigid wall and dummy reference particles are classified as *non-simplified* particles (fine particles marked with an "x" in the right-hand side of the border in Figure 2-b) while the fine resolution rigid wall and dummy particles at a distance of $2.0 \times l_{i,F}$ to the rigid wall and dummy reference particles are classified as *simplified* particles (highlighted fine particles in the right-hand side of the border in Figure 2-b). $l_{i,F}$ is the initial distance between particles of the fine resolution. Then, the next iteration adopts as *reference* particles the *simplified* particles obtained in the previous iteration and the process is repeated until all the near border fine resolution solid wall and dummy particles are classified and an equivalent coarse particle distribution of solid wall and dummy particles is achieved (Figure 2-c). Afterwards, the initial set of coarse resolution reference particles (Figure 2-a) plus the *simplified* fine resolution solid wall particles (Figure 2-c) are considered as *reference* particles for the simplification of the irregular distribution of the fluid particles (Figure 3).

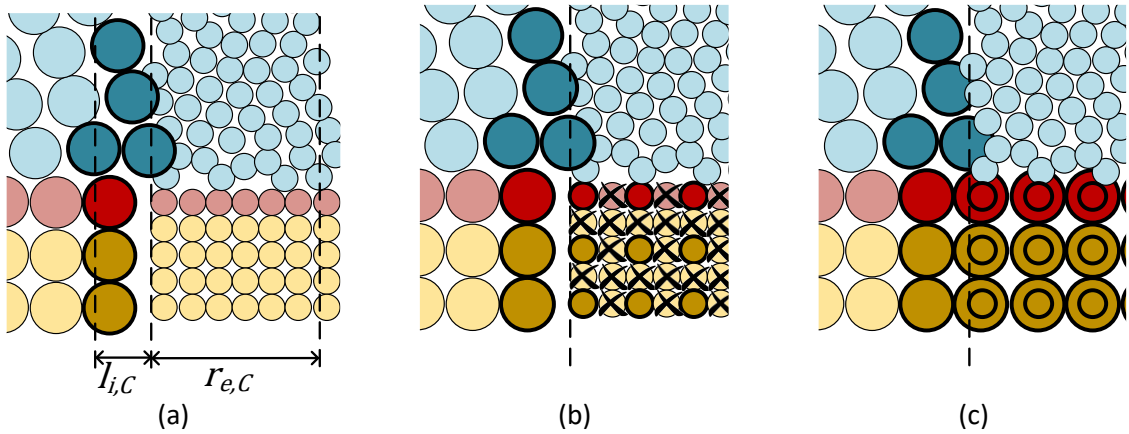

Figure 2 – Step-by-step scheme of the derefinement (simplification) algorithm to identify *simplified* and *non-simplified* fine resolution solid particles – fluid particle in blue, rigid wall particles in red and dummy particles in yellow; the resolution of the particles is indicated by its radius.

The simplification algorithm for the fluid particles has three main steps. In the first step, the fine resolution fluid particles which have at least three commonly shared reference particles within a distance of $1.6 \times l_{i,C}$ are considered as *candidate* particles (for instance, the set of fine resolution particles $\{a, \dots, h\}$ are candidate particles which share the same set of close reference particles $\{A, B, C\}$ in Figure 3-a). The criterion of three reference particles is adopted because the simplification should start at the corner between the initial set of coarse resolution reference particles and the solid wall *simplified* particles (region delimited by the thick dashed line in Figure 3-a).

In the second step, all the fine fluid particles within a distance of less than $0.8 \times l_{i,C}$ of any reference particle (Figure 3-b) are classified as *non-simplified* particles (fine resolution fluid particles marked with an "x" in Figure 3-c).

In the third step, an optimization criterion based on the distance between the *reference* particles and the remaining *candidate* particles defines the next *simplified* particle which yields the best approximation of an equivalent coarse resolution particle distribution. In this step, the distances between the *reference* particles and the remaining *candidate* particles are first calculated, see Figure 3-e to Figure 3-g. Next, the maximum distance to a reference particle is obtained for each candidate particle, e.g., the distances $d_{B,f}$, $d_{B,g}$ and $d_{B,h}$ in the examples presented in Figure 3-e,f,g. The next fine particle to be classified as a *simplified* particle is the one that has the shortest of such maximum distances, e.g., the *candidate* particle $h$ in the example of Figure 3, as $d_{B,h} < d_{B,g} < d_{B,f}$. The criterion of the distances can be written as:

$$\operatorname*{argmin}\left[\max_{i\in(\mathbb{I}_F\cup\mathbb{F}_F)}(d_{A,i}; d_{B,i}; d_{C,i})\right] \quad \to \quad i \text{ is } simplified \text{ particle} \tag{31}$$

where $\{d_{A,i}; d_{B,i}; d_{C,i}\}$ are the distances between the fine resolution candidate fluid particle $i \in (\mathbb{I}_F \cup \mathbb{F}_F)$, where subscript $F$ denotes fine particles. and each of the three references particles $\{A; B; C\}$.

Then, for the next iteration of the simplification process, the newfound *simplified* particle is included in the list of *reference* particles and the process is repeated, from the step in Figure 3-a to Figure 3-i, as many times as necessary until all the near-border fine resolution fluid particles are classified (Figure 3-j).

### 3.1.2 Refinement

The objective of the refinement is to obtain a background equivalent fine particle distribution from the position of the near-border coarse particles. In this way, the equivalent fine particle distribution is composed by the position of actual coarse particles, which are considered in this context as the position of fine resolution particles, and the position of the fine resolution *fictitious* particles that are created between the coarse particles. The refinement algorithm proposed herein is divided into three parts: the triangulation of the grid, the determination of the *fictitious* particles position and the interpolation of the physical quantities of the *fictious* particles.

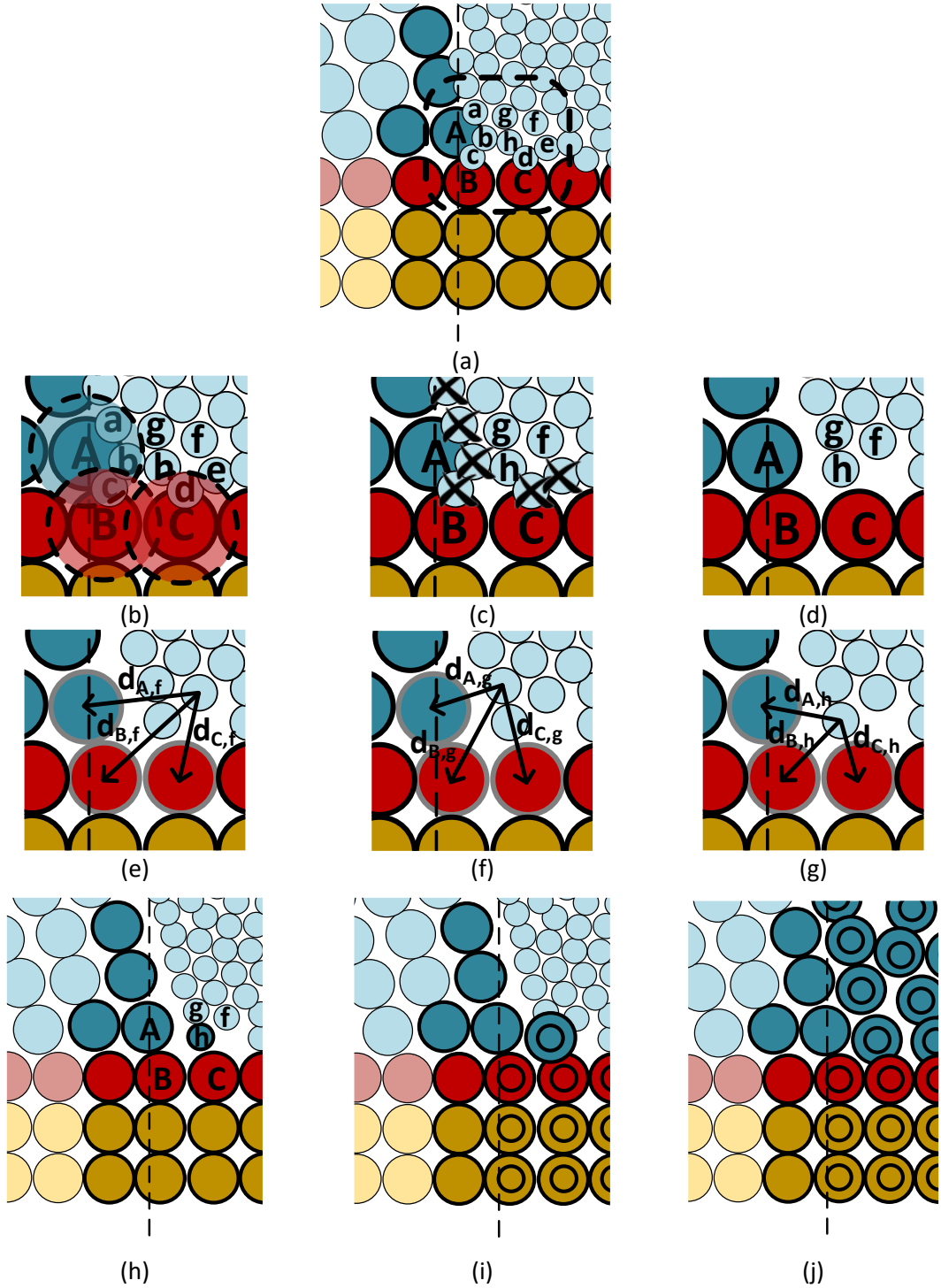

Figure 3 – Step-by-step scheme of the derefinement (simplification) algorithm to identify *simplified* and *non-simplified* fine resolution fluid particles – fluid particle in blue, rigid wall particles in red and dummy particles in yellow; the resolution of the particles is indicated by its radius.

The first step of the refinement technique is the triangulation process. It aims to obtain an initial estimation of the position of the *fictious* particles. The nodes of the grid to be triangulated are composed by the coarse resolution particles within a distance of $r_{e,C} = 3.1 \times l_{i,C}$ from the border (particles between the dashed lines at the left-hand side of the border in Figure 4-a,b) and the *simplified* fine resolution particles within a distance of $l_{i,C}$ of the border (highlighted particles between the dashed lines at the right-hand side of Figure 4-b). The nodes from the coarse resolution particles already indicate the position of a portion of the fine resolution particles in the background equivalent particle distribution. Then, the nodes of the grid closer than $1.6 \times l_{i,C}$ to each other are connected to by the lines shown in Figure 4-c. The nodes generated from *simplified* fine resolution particles are only connected to nodes from coarse particles, but not to other nodes from fine resolution *simplified* particles. On the other hand, nodes from coarse particles could connect to each other. The distance $1.6 \times l_{i,C}$ was defined empirically based on the particle distributions of several simulations.

Next, the initial location of the *fictitious* particles should be defined based on the connecting lines obtained in the previous step. In the case of lines that are not crossed by other line, a *fictitious* particle is placed exactly on the midpoint of the line. In the case of lines that are crossed by other line (such as in the intersecting lines at the region of the rigid wall particles and the fluid particle at the top-left in Figure 4-c), only a single *fictious* particle is generated by these two crossing lines. Hence, a list that links the $f$-th fictitious particle to the $k$-th line is created, in which a given *fictitious* particle could be associated to one or more lines but each line could only be associated to a single *fictitious* particle. Based on several tune-up MPS simulations, the authors observed that lines with no crossing and two crossing lines are by far the most common situations, while cases of more than two lines crossing each other are rare. For sake of simplicity, if a fictitious particle is created from two or more lines, its position is given by:

$$(x,y)_f = \left( \frac{\sum_k \alpha_{fk} \overline{x_{fk}}}{\sum_k \alpha_{fk}} ; \frac{\sum_k \alpha_{fk} \overline{y_{fk}}}{\sum_k \alpha_{fk}} \right) \qquad (32)$$

where $(\overline{x_{fk}}, \overline{y_{fk}})$ is the midpoint of the $k$-th line associated to the $f$-th *fictitious* particle and $\alpha_{fk}$ is a coefficient calculated for each line based on its length ($d_k$):

$$\alpha_{fk} = \frac{1}{2}\left[1 + \cos\left(\frac{2\pi}{1.6}\left(\frac{d_k}{l_{i,C}} - 0.8\right)\right)\right] \qquad (33)$$

where the constants 0.8 and 1.6 are related to the minimum and maximum distances of a connecting line. It weights the contribution of each line to the position of the associated *fictitious* particle, by giving more weight to the shorter lines ($\alpha_{fk} = 1$ for $d_k = 0.8 \times l_{i,C}$) and gradually decreasing the contribution to zero as the length of the line increases towards $d_k = 1.6 \times l_{i,C}$ ($\alpha_{fk} = 0$ for $d_k = 1.6 \times l_{i,C}$). As the coarse particles (nodes) evolve, this technique allows a smoother update of the initial position of the fictitious particles. This is a relevant issue for the stability of the computation because the initial position of a *fictitious* particle can change abruptly from one time step to the next time step as the topology of the coarse *reference* particles change, which might result in an abrupt change of equivalent particle distribution that induces some local pressure fluctuations. The background equivalent fine particle distribution is formed by the nodes (near-border coarse particles), considered as fine particles in this context, and the midpoints of the connecting lines (*fictitious* particles) (Figure 4-d).

As the physical quantities are not yet known in the position of the *fictious* particles, the last step of the refinement technique consists in the interpolation of the physical quantities at the centroid of the *fictitious* particles. At first, a particle type should be attributed to each *fictitious* particle. The material type of a given *fictitious* particle is attributed based on the *reference* particles associated to the nodes of the connecting lines used to create it. If one of the nodes is a fluid particle, the *fictitious* particle is a *fictitious* fluid particle. In the case one of the nodes is a dummy particle, the *fictitious* particle is a *fictitious* dummy particle. Finally, if all the nodes are solid wall particles, the *fictitious* particle is a *fictitious* wall particle. The result is a background equivalent fine resolution particle distribution such as the one presented in Figure 4-e.

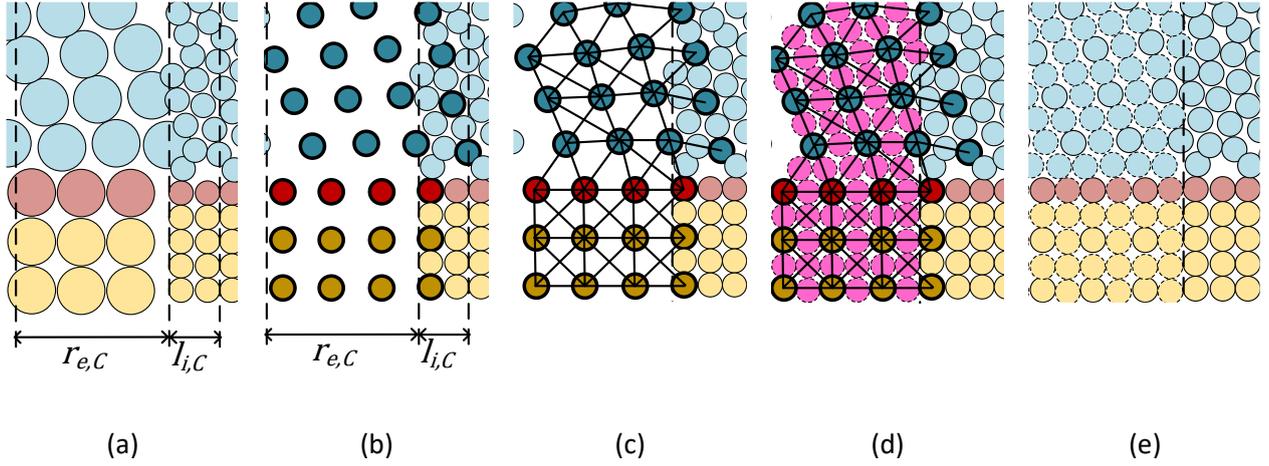

(a)      (b)      (c)      (d)      (e)

Figure 4 – Step-by-step scheme of the refinement algorithm to generate the initial position of the *fictitious* particles – fluid particles are in blue, rigid wall particles in red, dummy particles in yellow and *fictitious* particles in pink; the resolution of the particles is indicated by its radius.

Figure 5 shows the compact support adopted to obtain the velocity of the near-border $f_0$ fictitious particle, which considers the near-border coarse particles $\{A, B, C\}$ and the near-border fine particles $\{a, \dots, e\}$.

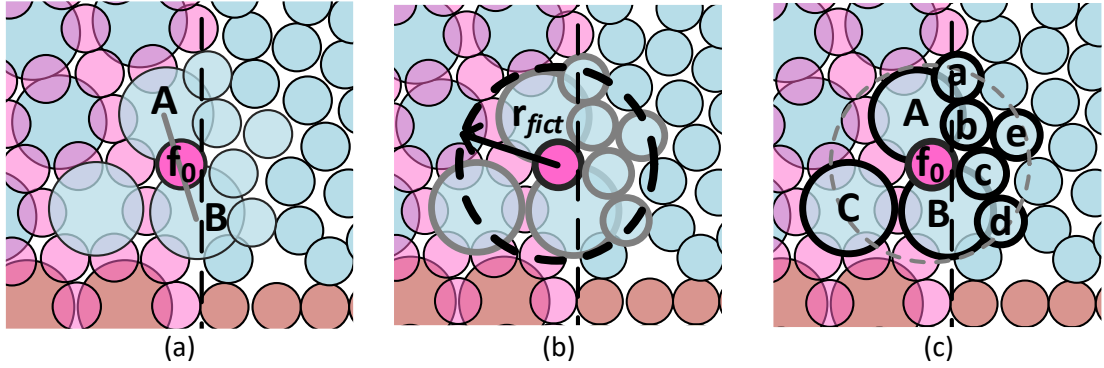

(a)      (b)      (c)

Figure 5 – Neighborhood of a *fictitious* particle for the interpolation of physical quantities – fluid particles are in blue, rigid wall particles in red and *fictitious* particles in pink; the resolution of the particles is indicated by its radius.

In the present study, the velocity vector at the center of a $f$-th *fictitious* particle is interpolated by:

$$\mathbf{u}_f = \frac{\sum_{g \in \mathbb{P}_f} \mathbf{u}_g}{N_{\mathbb{P}_f}} \tag{34}$$

where $\mathbb{P}_f$ represents only the $g$-th actual fine and coarse fluid particles used to obtain the $f$-th *fictitious* particle, with $N_{\mathbb{P}_f} = 2$ or $N_{\mathbb{P}_f} = 4$ being the number of actual particles.

### 3.1.3 Update of the border particles

At the end of the calculation, the particle distribution near the border should be updated as the fluid particles flow though the border between two subdomains. In the case the fluid flows from the coarse resolution towards the fine resolution subdomain, the coarse particles and the *fictitious* particles that cross the border become fine resolution particles (Figure 6). In the case the fluid flows from the fine resolution towards the coarse resolution subdomain, the *simplified* fine resolution particles that cross the border become coarse particles while the *non-simplified* fine resolution particles that cross the border are removed (Figure 7).

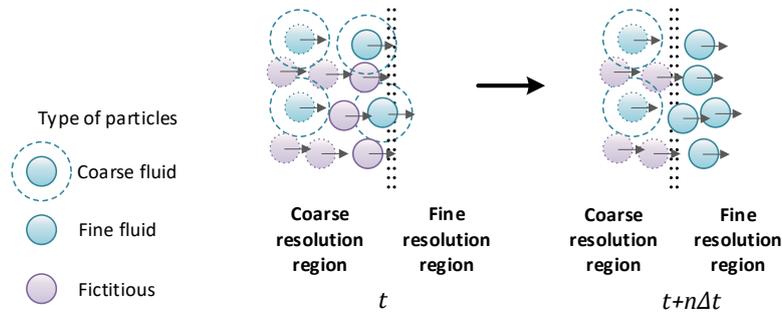

Figure 6 – Update of the particles in the border: flow from the coarse to the fine resolution subdomain.

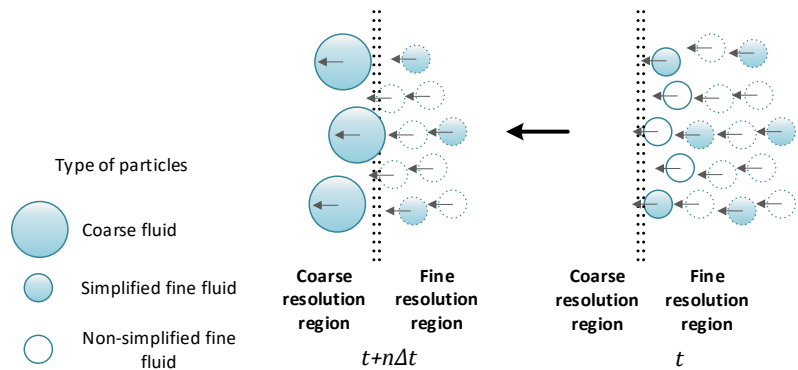

Figure 7 – Update of the particles in the border: flow from the fine to the coarse resolution subdomain.

## 3.2 Algorithm and pressure calculation

At the beginning of each time step, the velocity and position of all fluid particles are computed explicitly as described in Section 2.3, Eq. (19) to Eq. (24), for both the coarse and the fine subdomains. As mentioned before, for the inner fluid particles nearby the border of the subdomains, the contributions of the truncated neighborhood are computed using the background equivalent particle distributions obtained by refinement and derefinement (simplification) processes described in Section 3.1.

In a similar way, the update of the position and velocity by considering the explicit components and the collision are also applied to the *fictitious* particles within a distance of $r_{e,F} = 3.1 \times l_{i,F}$ from the border between subdomains in order to set properly the particle distribution for calculating the pressure of the near-border fine resolution particles. On the other hand, the position and velocity updates are not necessary for the *fictitious* particles with distance between $r_{e,F}$ and $2r_{e,F}$ from the border because they are outside of the neighborhood radius of the actual near border fine resolution particles and they are exclusively used as auxiliary neighbors to calculate explicit components of the *fictitious* particles within a distance of $r_{e,F}$ from the border.

### 3.2.1 PPE linear system

In the BMMR-MPS, the solution of the PPE must consider the *fictitious* and *simplified* particles within the neighborhood of the respectively actual fine and coarse particles near the border between subdomains. Let $b_i$ be the source term of PPE (right-rand terms) and $\Lambda_i$ be the fictitious neighborhood for a given $i$-th fine particle (see Figure 1-c), the first term in the left-hand side of Eq. (25) of PPE for a fine particle near the border can be rewritten as:

$$\frac{2dim}{\lambda^0 n^0}\left[\sum_{j\in\Omega_i}(P_j^{t+\Delta t} - P_i^{t+\Delta t})\omega_{ij} + \sum_{f\in\Lambda_i}(P_f^{t+\Delta t} - P_i^{t+\Delta t})\omega_{if}\right] - \frac{\rho}{\Delta t^2}\alpha_c P_i^{t+\Delta t} = b_i^*, \quad (35)$$

where $P_f$ designates the pressure at the $f$-th *fictitious* particle, here approximated by:

$$P_f = \frac{\sum_{g \in \mathbb{P}_f} P_g}{N_{\mathbb{P}_f}} \tag{36}$$

where $\mathbb{P}_f$ is the actual particles with pressure $P_g$ which was used as a node (i.e., *reference* particle) to create the $f$-th *fictitious* particle in the refinement (see Section 3.1.2), and $N_{\mathbb{P}_f} = 2$ or $N_{\mathbb{P}_f} = 4$ designates the number of actual particles. The actual particles $g$ can be any of the inner particles $i$ and $j$ in Eq. (35), and $P_f$ is given by a function $F$ of the variables $P_g$, i.e., $P_f = F(P_g)$. Substituting Eq. (36) in Eq. (35), $P_f = F(P_g)$ are not considered as unknown variables in the linear PPE system. Thus, the matrix of coefficients **A** becomes asymmetrical for the linear PPE system.

Concerning a target $i$-th coarse resolution particle near the border, the first term in the left-hand side of Eq. (25) of PPE reads:

$$\frac{2dim}{\lambda^0 n^0} \left[ \sum_{j \in (\Omega_i - \Psi_i)} (P_j^{t+\Delta t} - P_i^{t+\Delta t}) \omega_{ij} + \sum_{s \in \Psi_i} (P_s^{t+\Delta t} - P_i^{t+\Delta t}) \omega_{is} \right] - \frac{\rho}{\Delta t^2} \alpha_c P_i^{t+\Delta t} = b_i^*, \tag{37}$$

where $\Psi_i$ represents the *simplified* particles neighborhood of the $i$-th actual coarse resolution particle (orange particles in Figure 1-b) and pressure $P_s$ represents the pressure at the $s$-th *simplified* particle, which is a known variable in the linear PPE system because it is the value of its associated fine particle. It is important to point out that the $s$-th *simplified* particle is neighbor of particle $i$ but the particle $i$ is not necessarily neighbor of particle $s$, i.e., Eq. (37) is not necessarily symmetrical between particles $i$ and $s$, and consequently, this also leads to an asymmetrical matrix of coefficients **A**.

In summary, the final form of **A** is a bandwidth but asymmetrical matrix for the solution of the linear PPE system in the proposed BMMR-MPS. To clarify the configuration of the matrix of coefficients **A** of the PPE, the matrices in the single-resolution and multi-resolution MPS for a hydrostatic tank are presented in Figure 8. The colored points are related to the non-zero elements of the matrix of coefficients.

The PPE linear system in the original single-resolution MPS method has a sparse matrix **A** in which the non-zero elements are concentrated in its diagonal, as illustrated in Figure 8-a. On the other hand, non-zero elements that relate particles with different resolutions are observed outside the main diagonal in the resultant matrix **A** of the BMMR-MPS, as presented in Figure 8-b. The smaller portion of non-zero elements near the bottom-right corner (①) are related to the coarse resolution particles and the larger portion of non-zero elements in the top-left corner (②) are related to the fine resolution particles. The non-zero elements related to the *simplified* neighborhood of the coarse resolution particles are in the bottom-left corner (③) while the non-zero elements related to the *fictitious* neighborhood of the fine resolution particles are in the top-right corner (④).

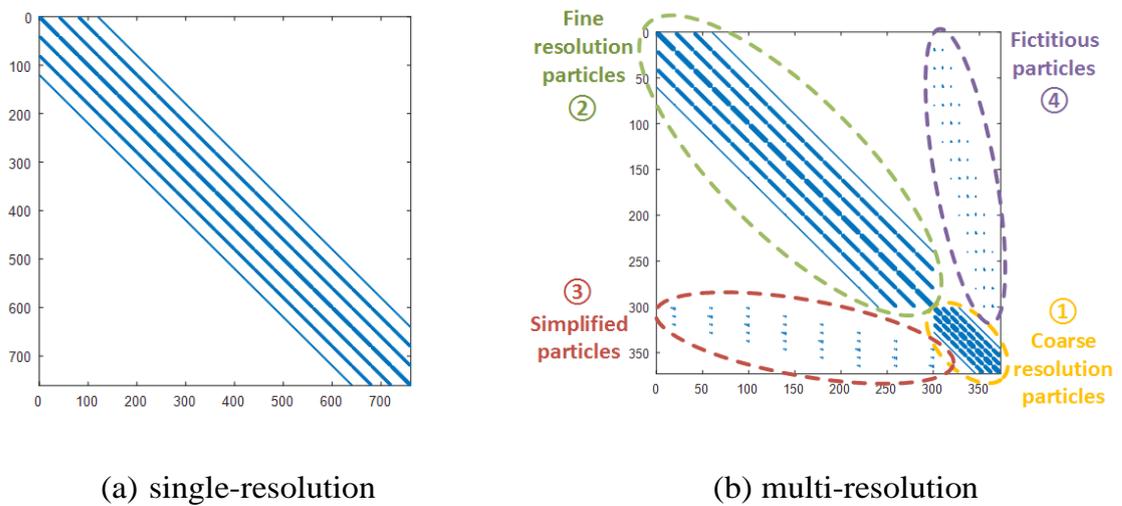

(a) single-resolution        (b) multi-resolution

Figure 8 - Non-zero elements of the PPE matrix of coefficients **A** for (a) single and (b) multi-resolution MPS.

With certain boundary conditions, the sparse linear system of PPE may be efficiently solved by using iterative methods, such as the conjugate gradient (CG) in the diagonal and symmetrical single-resolution matrix, while the generalized minimum residual (GMRES) method can be used to solve the asymmetrical multi-resolution matrix. A comparison between the computational time to solve single and multi-resolution PPE is evaluated in Section 5.

### 3.2.2 Source term near the border

During the update of the particles in the border, slightly discontinuities on particle number density may occur due to approximated particle distribution. Since the source terms based on particle number density deviation are very sensitive to the particle positions, which may lead to more unstable pressure calculation, we adopted only the velocity-divergence-free condition as the source term of PPE for particles within a distance of $2.0 \times l_{i,C}$ from the border between subdomains to further smooth the transition between subdomains:

$$\begin{cases} \langle \nabla^2 P \rangle_i^{t+\Delta t} - \dfrac{\rho}{\Delta t^2} \alpha_c P_i^{t+\Delta t} = c_s \dfrac{\rho}{l_i} \langle \nabla \cdot \mathbf{u} \rangle_i^{**} & \|\mathbf{r}_i - \mathbf{r}_b\| \leq 2 \times l_{i,C} \\ \langle \nabla^2 P \rangle_i^{t+\Delta t} - \dfrac{\rho}{\Delta t^2} \alpha_c P_i^{t+\Delta t} = c_s^2 \dfrac{\rho}{l_i^2} \left( \dfrac{n^0 - n_i^{**}}{n^0} \right) + c_s \dfrac{\rho}{l_i} \langle \nabla \cdot \mathbf{u} \rangle_i^{**} & \text{otherwise} \end{cases} \quad (38)$$

where the value of $\|\mathbf{r}_i - \mathbf{r}_b\|$ represents the distance between the particle $i$ from the border between subdomains, where $\mathbf{r}_b$ is the closest point on the border to particle $i$, and $l_{i,C}$ is the initial distance between particles of the coarse resolution. Through consideration of Eq. (38), the source term near for near-border particles is illustrated in Figure 9.

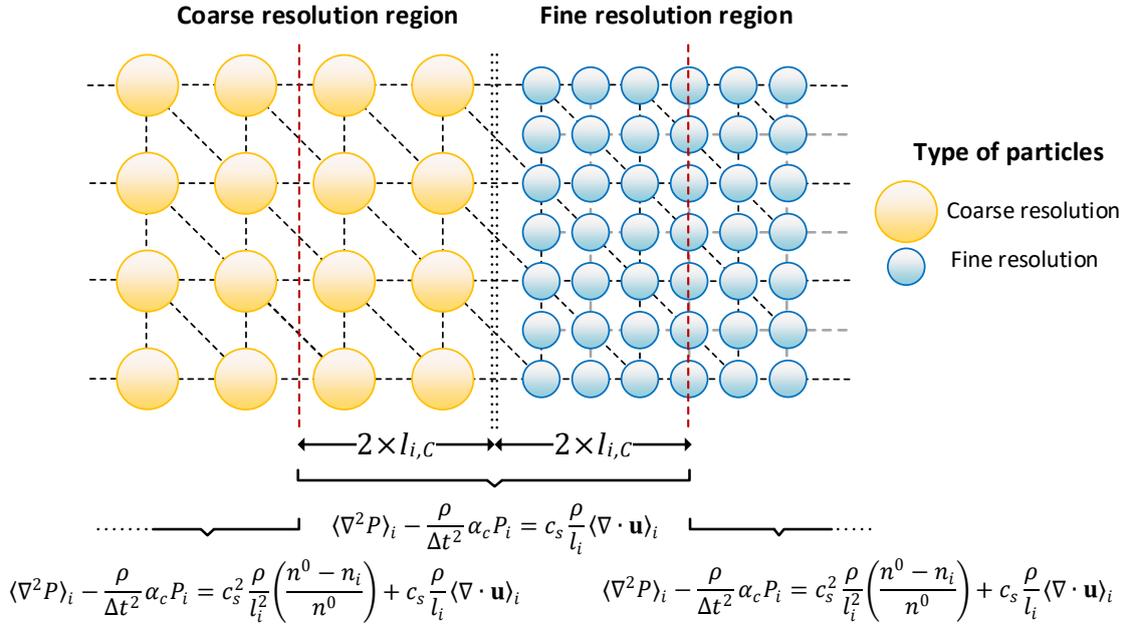

Figure 9 – Source term of the PPE for particles near the border.

### 3.2.3 Algorithm of the BMMR-MPS

The overall solution procedure of the BMMR-MPS method is detailed in Figure 10:

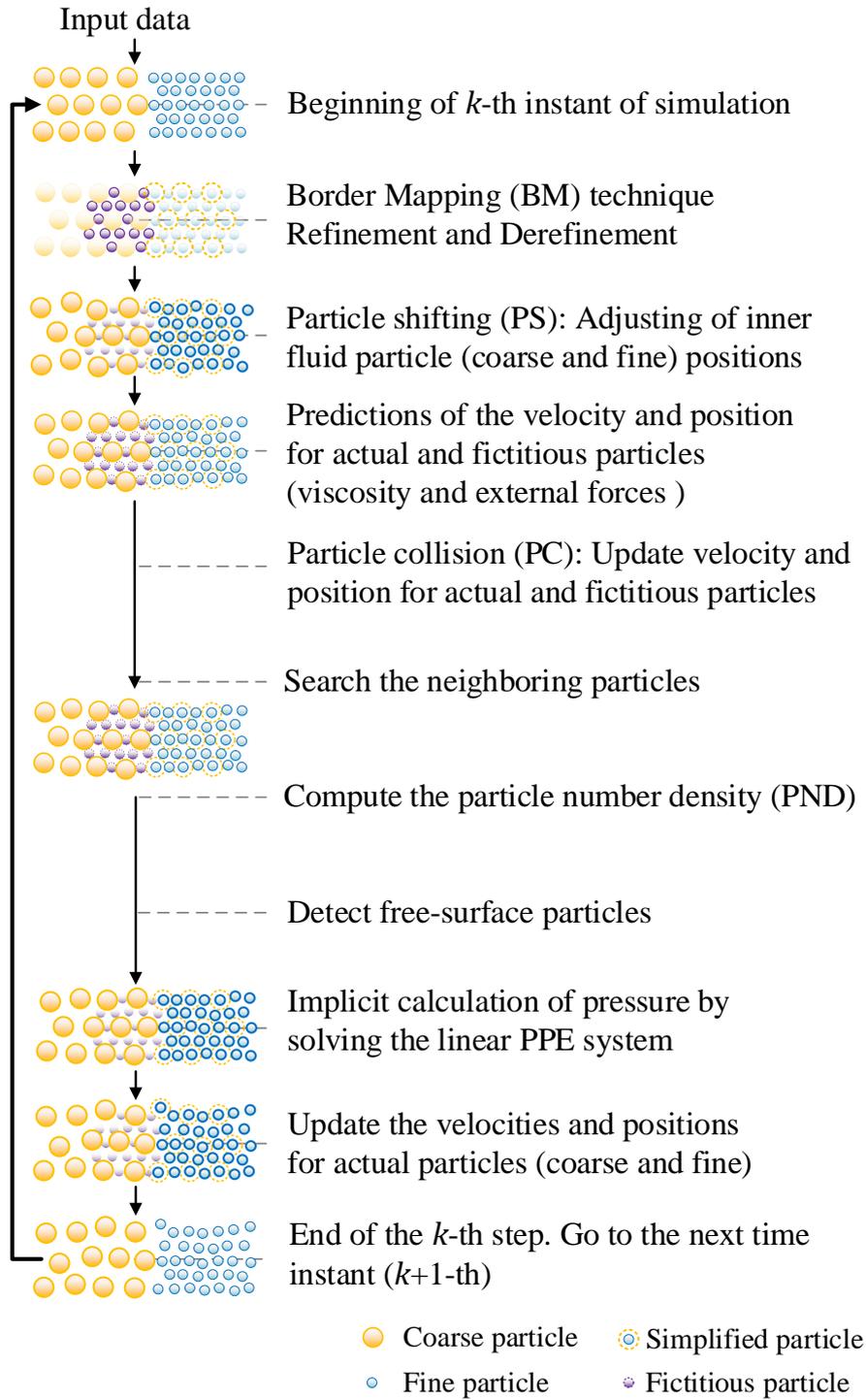

Figure 10 – Flowchart of the BMMR-MPS method.

### 3.2.4 Comments on the BMMR-MPS and other multi-resolution techniques

The BMMR-MPS adopts closed polygon refinement zones predefined by the users and maintains the resolution within each subdomain by refinement and derefinement (simplification) algorithms that avoid abrupt changes in the particle distribution in the border.

Using the split-merge terminology for clarity, the BMMR-MPS refinement algorithm uses a triangulation to create a *daughter* particle in the exact position of the *mother* particle and creates the *fictitious* particles in a better suited position in relation to the local particle distribution, while the simplification algorithm creates the *mother* particles in the exact position of one of its *daughter* particles. In this way, the abrupt local fluctuations of the particle density are avoided. Such assertion is indirectly supported by the findings from Vacondio et al. (2016), which concluded that the density error that results from different patterns of splitting is minimized when the pattern has a *daughter* particle that is created at the same position of the *mother* particle. Besides, by using pre-defined *daughter* particle patterns, most of the splitting algorithms in literature (see Table 1) only consider the position of a single *mother* particle to create its *daughter* particles but effectively disregard the distribution of the neighbor particles. On the other hand, using the BMMR-MPS, the *fictitious* particles are not created from a single *mother* particle, but from the relative position between two or more *mother* particles, which result in a more consistent overall particle distribution.

The ratios of the refinement and the derefinement are the same and only particles with the same resolution are refined or simplified so that each subdomain is strictly kept single-resolution during the entire simulation without disordered mixing of particles of different resolutions. Meanwhile, the majority of other multi-resolution techniques in literature adopt different ratios of refinement and derefinement or allow particles of different sizes to merge, which result in a gradual and steady creation of particles of different sizes within each subdomain as the simulation evolves. In such context, the BMMR-MPS scheme is robust enough to provide stable computations for highly irregular particle distributions using only the original single-resolution framework of the MPS method and allows the adoption of the additional treatments commonly used to improve the stability and accuracy of the simulations. Actually, by means of the BMMR technique, the application of the original single-resolution formulations to the multi-resolution one is quite straight-forward because no substantial changes are required to the discrete differential operators and boundary conditions. Moreover, BMMR-MPS has a strong coupling between subdomains of different resolutions, which is very important from the viewpoint of the solution of the PPE system.

Concerning the Newton's third law for interparticle forces, the proposed method does not strictly ensure it in the border region. Shibata et al. (2012; 2017), Hu et al. (2020) and Yang et al. (2021) adopt similar approaches in which the neighborhood radius (or smoothing length) of each particle is associated to its own size/resolution. The asymmetry of the matrix of coefficients of the PPE is a consequence of that. As advantages of such approach, the neighborhood radius is proportional to the resolution of the particles and its compact support is always symmetrical, i.e., circular in the 2D case. Besides, all the particles have almost the same number of neighbors, which is advantageous from a computational standpoint by making it easier to allocate the memory for the neighbor particle list. In the literature, one way to enforce the Newton's third law for interparticle forces is to use a single neighborhood radius for the entire computational domain (Chen, et al., 2016), which impose the use of an unnecessarily large neighborhood radius in fine resolution regions. Other way to ensure the Newton's third law for interparticle forces is to use an averaged neighborhood radius for each pair of neighbor particles (Vacondio, et al., 2013; Tanaka, et al., 2018), which results in an asymmetrical compact support for the interpolation of physical quantities in the border between subdomains.

## 4   Numerical examples

In order to verify the accuracy of the proposed multi-resolution MPS, three 2D benchmark free-surface flows are considered. First, an inviscid standing wave is simulated and the computed wave amplitude is compared with the $2^{nd}$ order analytical solution along with time. Moreover, the mechanical energy dissipation is also verified. Second, a simulation of a dam-break flow is performed and pressure and free-surface profile are compared with the experimental data. Finally, the water impact of a rigid circular cylinder is simulated to show the capability of the present BMMR-MPS to tackle highly nonlinear FSI problems. The computed free-surface profile and cylinder penetration are compared against experimental data.

## 4.1 Inviscid standing wave: conservation of mechanical energy

To illustrate the accuracy and conservation properties (volume and energy) of the proposed BMMR-MPS, the evolution of a 2D inviscid standing wave, widely studied in particle methods (Suzuki, et al., 2007; Antuono, et al., 2011; Souto-Iglesias, et al., 2013; Khayyer, et al., 2017), was simulated. Figure 11 shows the main dimensions of the standing wave of mean water level $H_s = 1$m and length $\lambda_s = 2$m. For the multi-resolution simulation, the initial particle distance $l_c = 0.01$m is adopted in the coarse subdomain defined in the subdomain below the height $H_c = 0.6$m, whereas $l_f = 0.005$m is used in the fine subdomain in the remaining upper subdomain. For the purpose of comparison, numerical models with single and uniform resolutions $l_i = 0.01$ and 0.005m were also simulated.

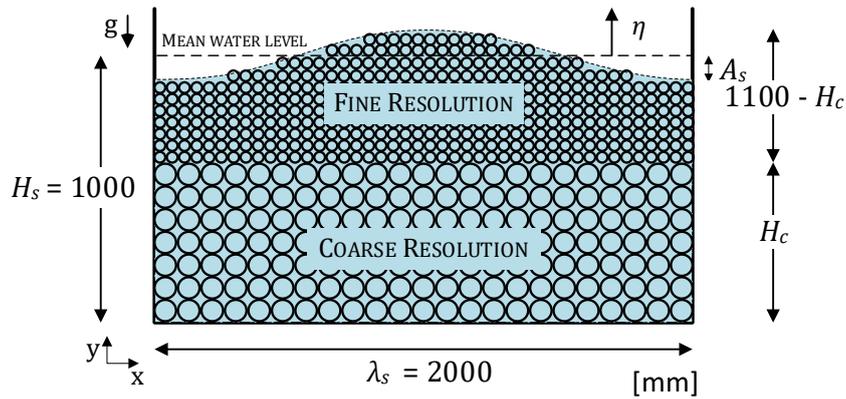

Figure 11 - Schematic view of inviscid standing wave. Geometry and main dimensions of the initial condition. The sizes of the particles are only for illustration purpose and do not correspond to the actual values.

The initial surface elevation (adopting the zero at the mean water level) is given by:

$$\eta_0(x) = A_s \cos[k_2(x + \lambda_s/2)], \tag{39}$$

where $A_s = 0.1$m denotes the wave amplitude and the constant $k_2 = 2\pi/\lambda_s$ is the wave number. The fluid of density $\rho = 1000$kg/m³ is assumed inviscid ($\nu = 0.0$). The gravity is $g = 9.81$m/s². The numerical parameters are summarized in Table 2.

As a reference, the analytical solution $\eta(t)$ provided by Wu and Taylor (1994) is adopted:

$$\eta(t) = \eta_{1st}(t) + \eta_{2nd}(t), \tag{40}$$

$$\eta_{1st}(t) = A_s \cos(\omega_2 t), \tag{41}$$

$$\eta_{2nd}(t) = \frac{1}{8g}\left\{2(\omega_2 A_s)^2 \cos(2\omega_2 t) + \frac{A_s^2}{\omega_2^2}[k_2^2 g_2^2 + \omega_2^4 - (k_2^2 g^2 + 3\omega_2^4)\cos(\omega_4 t)]\right\}, \tag{42}$$

with

$$k_m = \frac{m\pi}{\lambda_s}; \quad \omega_m = \sqrt{k_m g \tanh(k_m H_s)}. \tag{43}$$

Table 2 - Inviscid standing wave. Numerical parameters.

| Parameter | Value | Parameter | Value |
|---|---|---|---|
| Particle distance $l_i$ (m) | Coarse 0.02<br>Fine 0.01 | Collision distance $\alpha_1$ | 0.85 |
| Effective radius $r_e$ (m) | Small 2.1×$l_i$<br>Large 3.1×$l_i$ | Coefficient of restitution $\alpha_2$ | 0.2 |
| Time step $\Delta t$ (s) | 5.0×10$^{-4}$ | Surface threshold $\beta_F$ | 0.93 |
| Speed of perturbations $c_s$ (m/s) | 2.0 | Surface threshold $\varrho_F$ | 0.25 |
| Artificial compressibility $\alpha_c$ (ms²/kg) | 2.0×10$^{-8}$ | | |

Figure 12 depicts the evolution of the standing wave and its computed pressure fields. The fundamental period is given by $T = \frac{2\pi}{\omega_2} \cong 1.134$s. The free-surface evolution is well reproduced by the single and multi-resolutions simulations. Furthermore, a continuous and smooth pressure field is computed by the proposed BMMR-MPS simulations at all selected instants.

Long-time evolution of water surface elevations amid the rectangular tank ($x = 1.0$m) are illustrated in Figure 13(a). All numerical simulations reproduced the wave heights in very good agreement with the analytical one, although slight numerical damping is present. These results indicate that conservation features (volume and mechanical energy) along reasonable long-time simulations do not deteriorate by BMMR-MPS.

The mechanical energy ($E_M$), including kinetic and potential energy, is computed by:

$$E_M = \rho \left( \sum_{i \in \mathbb{F} \cup \mathbb{I}} l_i^{dim} \frac{\|\mathbf{u}_i\|^2}{2} - \sum_{i \in \mathbb{F} \cup \mathbb{I}} l_i^{dim} \mathbf{g} \cdot \mathbf{r}_i \right). \qquad (44)$$

where $\mathbf{r}_i$ denotes the position vector from the tank bottom to the particle $i$.

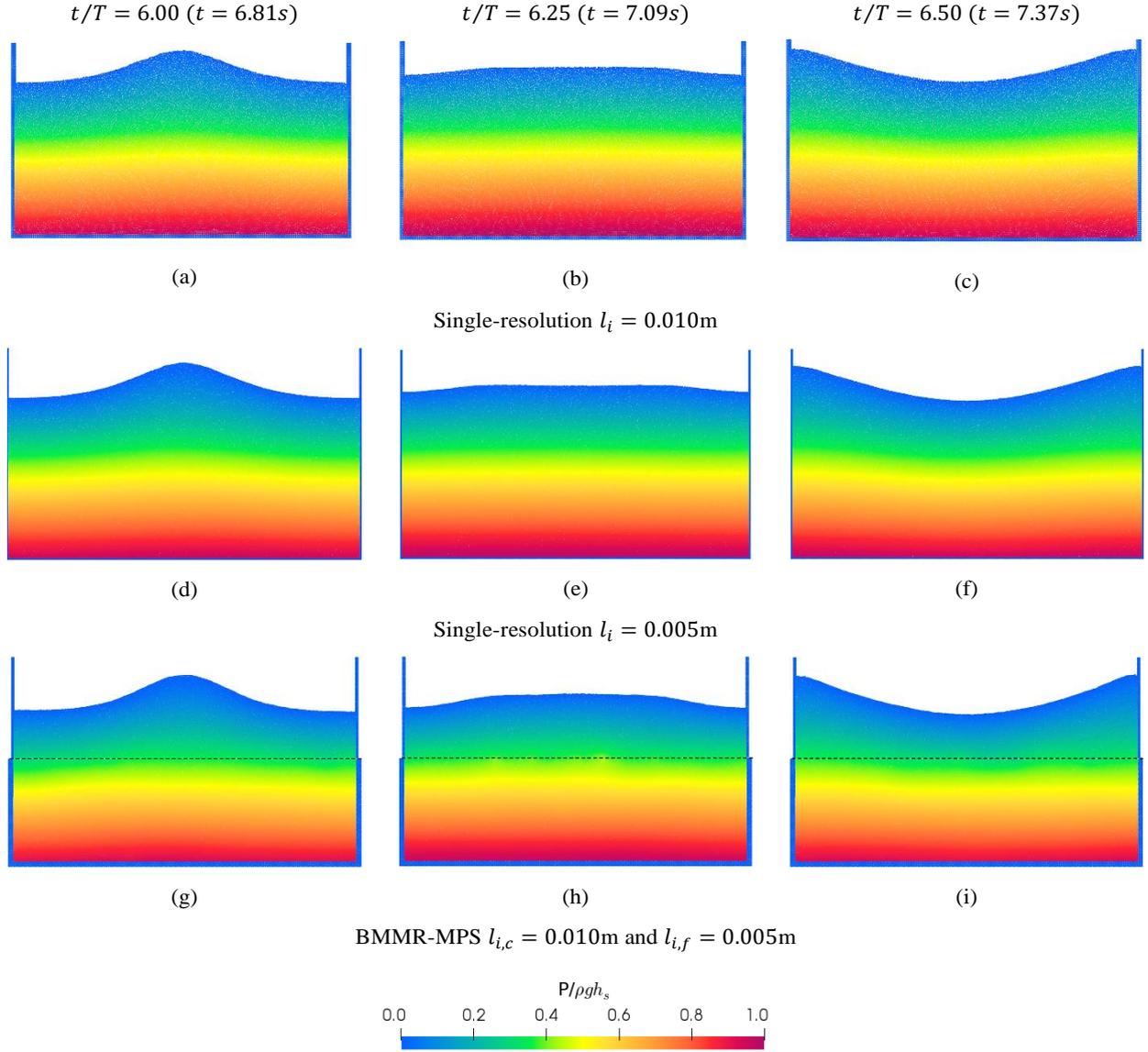

Figure 12 - Evolution of the inviscid standing waves computed at the instants $t/T = 6.00$, $6.25$ and $6.50$ ($t = 6.81$, $7.09$ and $7.37$s). The colors of the fluid particles are related to its pressure and the black dashed line illustrates the division between the subdomains.

Figure 13(b) shows that the computed mechanical energy variation $(E_M/E_{Mo} - 1)$ using uniform and multi-resolutions are respectively within 1.5% and 4.0% of the initial mechanical energy ($E_{Mo}$). These results indicate that the amount of mechanical energy dissipated using uniform and BMMR-MPS are almost of the same order, although multi-resolution has a slightly larger numerical dissipation.

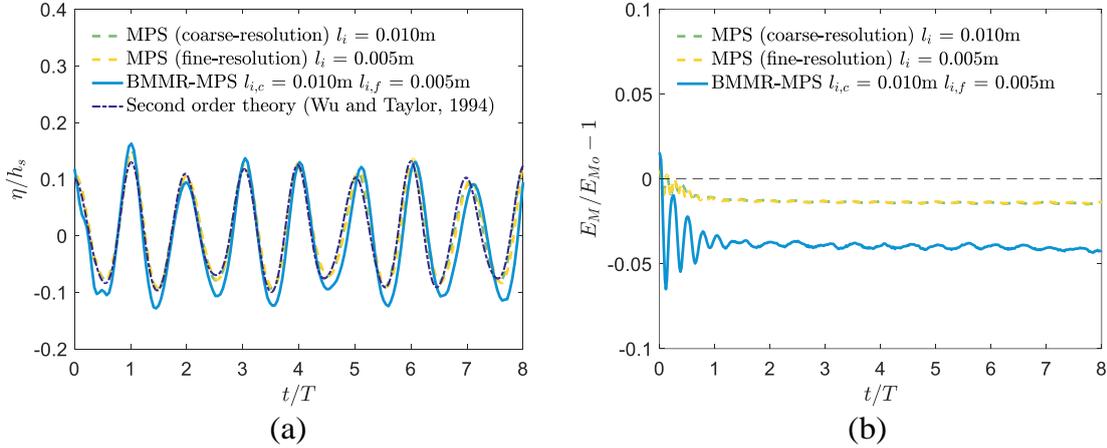

Figure 13 - Time histories of the water surface elevation at the center of the tank ($x = 1.0$m) from the analytical solution (Wu & Taylor, 1994) and present results. Simulations using single-resolution MPS with particle distances $l_i = 0.010$m (coarse-resolution) and $0.005$m (fine-resolution), and BMMR-MPS $l_{i,c} = 0.010$m, $l_{i,f} = 0.005$m. (b) Deviation of mechanical energy of the numerical simulations. The subscript '0' indicates the analytical value of the mechanical energy.

## 4.2 Dam breaking: mass conservation and hydrodynamic pressure

In this section, we simulated the evolution of dam-break flow, which has been widely used to verify the accuracy of particle-based methods in free-surface flows (Koh, et al., 2012; Barcarolo, et al., 2014; Tanaka, et al., 2018; Jandaghian & Shakibaeinia, 2020). Figure 14 displays the initial geometry of the problem (Lobovský, et al., 2014), a rectangular tank of height $H_T = 0.6$m, length $L_T = 1.61$m, initial water column of height $H_w = 0.3$m and length $L_w = 0.6$m. The pressure is monitored at the sensor SD$_1$ placed at the right wall, 3mm from the bottom of the tank. The fluid properties are density $\rho = 997$ kg/m$^3$ and kinematic viscosity $\nu = 8.9 \times 10^{-7}$m$^2$/s, and the value of gravity $g = 9.81$m/s². Table 3 highlights the numerical parameters.

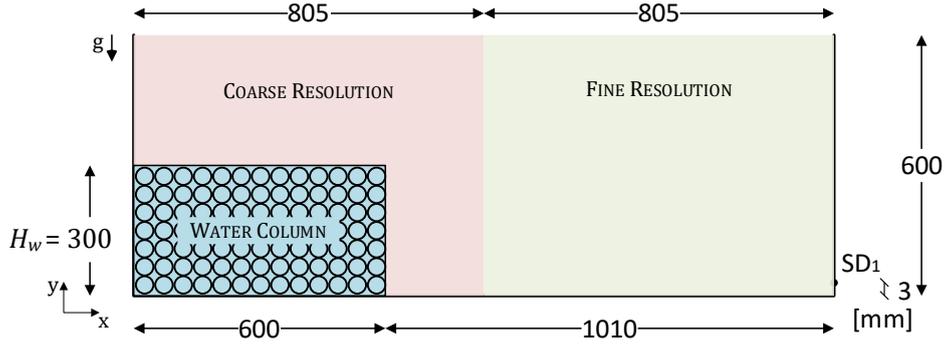

Figure 14 - Main dimensions of the initial condition of the dam breaking of initial water column $H_w = 0.3$m and pressure sensor position (SD$_1$) (Lobovský, et al., 2014). The sizes of the particles are only for illustration purpose and do not correspond to the actual values.

Table 3 - Dam breaking. Numerical parameters.

| Parameter | Value | Parameter | Value |
| --- | --- | --- | --- |
| Particle distance $l_i$ (m) | Coarse 0.05<br>Fine 0.0025 | Collision distance $\alpha_1$ | 0.85 |
| Effective radius $r_e$ (m) | Small $2.1 \times l_i$<br>Large $3.1 \times l_i$ | Coefficient of restitution $\alpha_2$ | 0.2 |
| Time step $\Delta t$ (s) | $2.5 \times 10^{-4}$ | Surface threshold $\beta_F$ | 0.93 |
| Speed of perturbations $c_s$ (m/s) | 2.0 | Surface threshold $\varrho_F$ | 0.25 |
| Artificial compressibility $\alpha_c$ (ms²/kg) | $1.0 \times 10^{-8}$ | | |

Figure 15 presents the experimental and numerical dam-break flow evolutions at the non-dimensional instants $t(g/H_w)^{1/2} =$ 3.27, 5.85 and 6.67 ($t =$ 0.573, 1.023 and 1.167s). The colors on the fluid particles are related to its non-dimensional pressure $P/\rho g H_w$. After it reaches the downstream wall, portion of the flow forms a vertical runup jet at $t(g/H_w)^{1/2} = 3.27$. As the dam-break flow proceeds, the vertical jet descends under the gravity force and a backward wave is generated at $t(g/H_w)^{1/2} = 5.85$. Afterwards, the backward wave collapses as a plunging wave at $t(g/H_w)^{1/2} = 6.67$. The computed wave profiles using single and multi-resolution are in reasonable agreement with the experimental one. Moreover, smooth and continuous pressure distribution was computed by BMMR-MPS, even across the multi-resolution border.

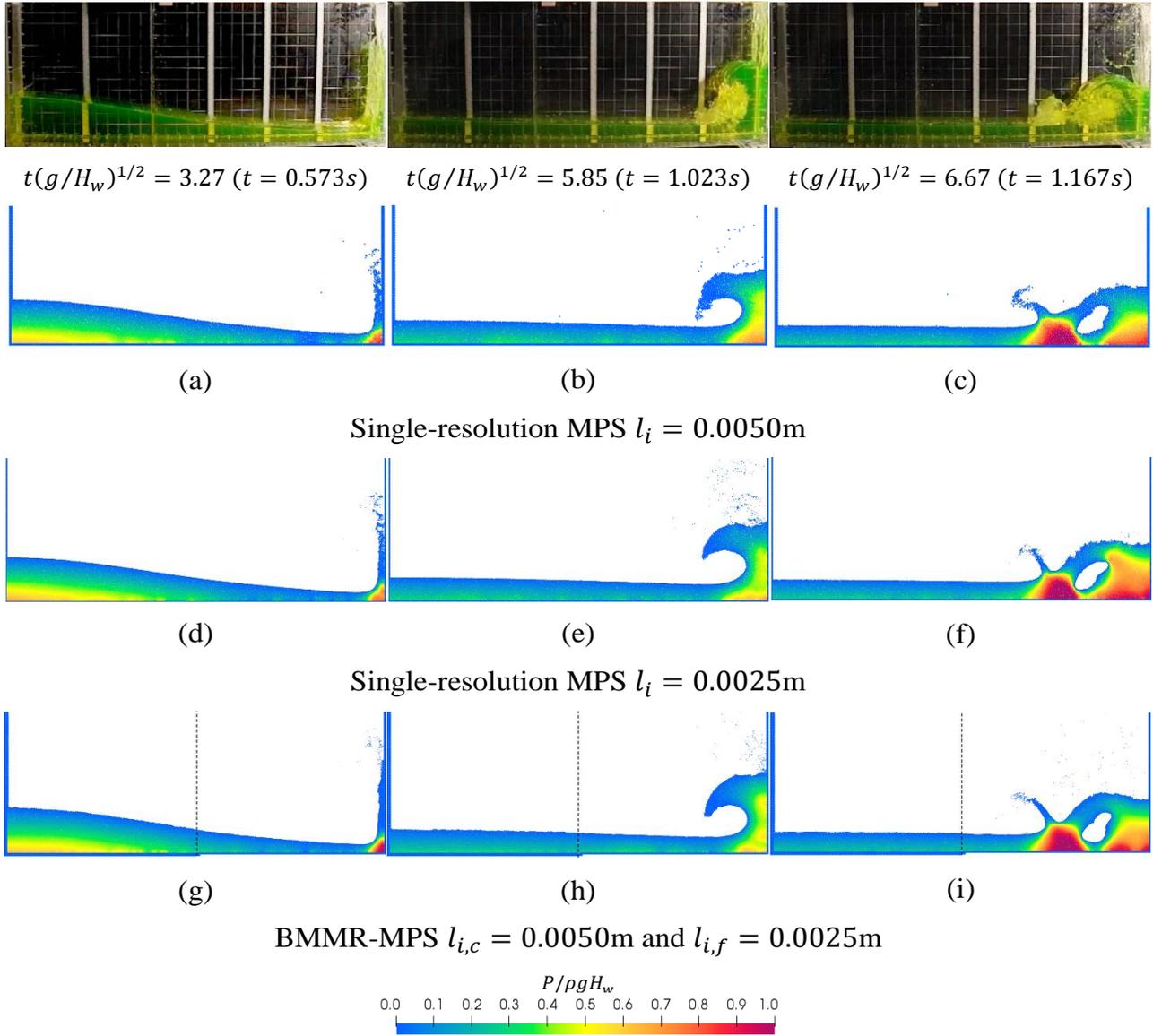

Figure 15 - Dam breaking evolution from the experiment (Lobovský, et al., 2014) and numerical simulations at the instants $t(g/H_w)^{1/2} = 3.27$, 5.85 and 6.67 ($t = 0.573, 1.023$ and $1.167s$). The colors of the fluid particles are related to its pressure.

In order to show the particle distribution consistency near the border, Figure 16 displays the positions of actual, *simplified* and *fictitious* particles at the instants $t(g/H_w)^{1/2} = 3.27$ and $= 5.85$ obtained with BMMR-MPS. Coarse and fine domains are represented by yellow and blue particles, respectively. As shown in Figure 16, the equivalent coarse-resolution (*simplified* green particles) obtained by the derefinement process, see Section 3.1.1, presents a particle arrangement very close to the actual coarse particle distribution. At the same time, the equivalent fine-resolution (*fictitious* purple particles) obtained by the refinement process, see Section 3.1.2, is very similar to that given by the actual fine particle positions.

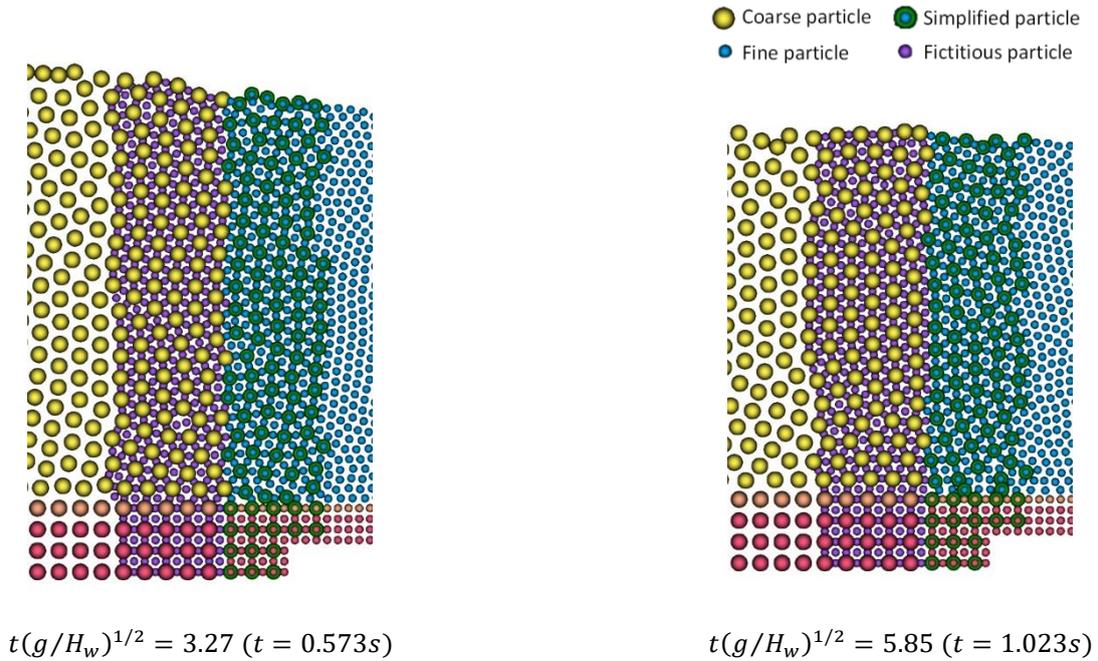

$t(g/H_w)^{1/2} = 3.27$ ($t = 0.573s$)　　　　　　$t(g/H_w)^{1/2} = 5.85$ ($t = 1.023s$)

Figure 16 – Distribution of coarse, fine, *fictitious* and *simplified* particles near the border.

The quantitative analysis of the mass conservation in the BMMR-MPS is provided by Figure 17(a), which presents the time series of the total area of the particles, proportional to the mass in the case of the 2D simulation. Only the fluid particles were considered. The values of total mass (coarse + fine resolutions) along the time are also plotted. The decrease rate of the mass of coarse resolution particles is almost equal to the increase rate of fine resolution particles. For this highly nonlinear hydrodynamics case, the proposed technique leads to a satisfactory result of only 1.0% loss of total mass at $t = 10.0$s.

Figure 17(b) provides the raw pressure time histories measured in the experiment (Lobovský, et al., 2014) and numerically computed at the sensor $SD_1$. The shadowed area represents the lower and upper bounds of 2.5% and 97.5% estimation in the experiments. The instant of the first peak pressure computed with fine single resolution and proposed multi-resolution agree well with the experiment, whereas a delay occurs for the coarse resolution. Concerning the pressure magnitude, the computed results are in good agreement with the experimental one, and unphysical pressure oscillations are reasonably low. Besides that, the magnitude of the computed pressure peaks using the multi-resolution is close to the median experimental values whereas by using single resolutions they are close to the upper bound estimated by the experiments. One should be remarked that the computed impact pressure peaks are considerably sensitive and prone to variations with respect to the numerical parameters. In summary, these results demonstrated that the BMMR-MPS reproduces well the dam break evolution and ensures an accurate and smooth computed pressure.

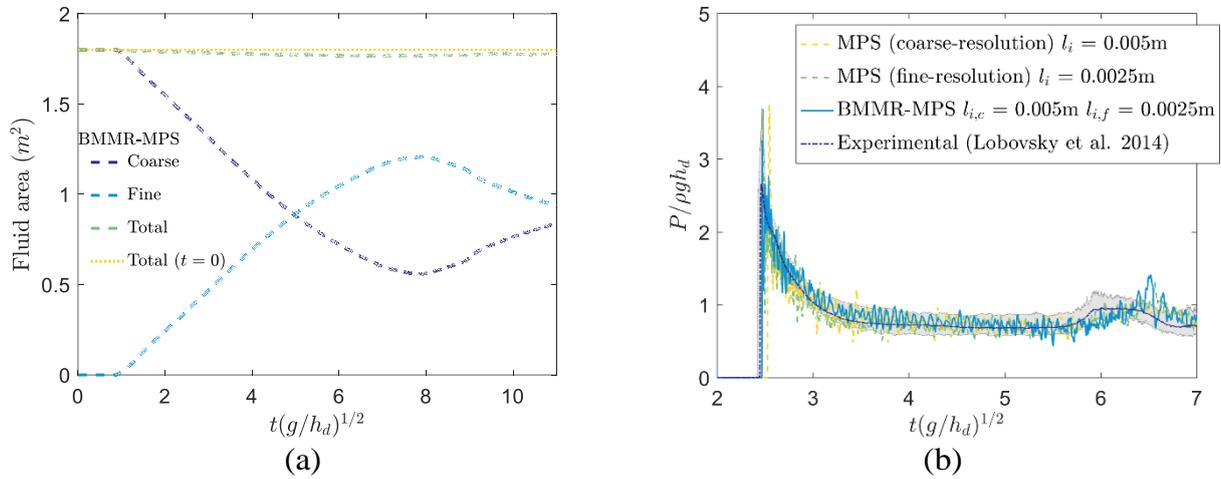

Figure 17 – (a) Fluid area over time. (b) Raw pressure time histories at the sensor $SD_1$ from the dam breaking experiment (Lobovský, et al., 2014) and the present results using single-resolution MPS and multi-resolution BMMR-MPS. The gray region shows the lower and upper bounds of 2.5% and 97.5% estimation in the experiments.

### 4.3 Water impact of half buoyant circular cylinder: fluid-structure interaction

In the present section we consider the impact of a half buoyant circular cylinder, i.e., a cylinder of density equal to half the fluid density, which has been experimentally studied by Greenhow and Lin (1983). In the experiment, the cylinder with diameter $D_s = 0.11$m is initially located 0.5m above the water surface and then starts to move in free fall reaching the initial impact velocity of 2.955m/s. In the numerical modeling, we considered only half of the cylinder, with free-slip boundary condition imposed in the symmetry plane, as shown in Figure 18. The fluid domain consists of a rectangular tank of length $L_T = 1.0$m and an initial water column of height $H_w = 0.3$m. The cylinder's bottom is positioned $2 \times l_i$ above the water surface at the beginning of the simulation, with an initial downward vertical velocity $v_s = 2.955$ m/s. The fluid properties are density $\rho = 1000$ kg/m$^3$ and kinematic viscosity $\nu = 1 \times 10^{-6}$ m$^2$/s, and the value of gravity $g = 9.81$m/s². The density of the half buoyant cylinder is $\rho_s = 500$ kg/m$^{-3}$. Table 4 gives the numerical parameters.

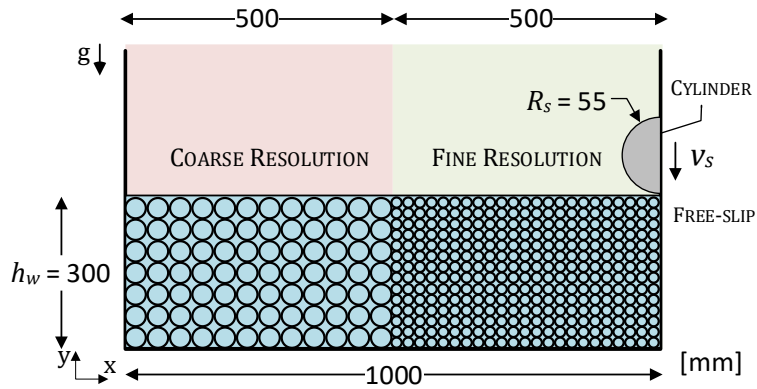

Figure 18 - Schematic view of water impact of half buoyant circular cylinder. Geometry and main dimensions of the initial condition. The sizes of the particles are only for illustration purpose and do not correspond to the actual values.

Table 4 - Water impact of half buoyant circular cylinder. Numerical parameters.

| Parameter | Value | Parameter | Value |
| --- | --- | --- | --- |
| Particle distance $l_i$ (m) | Coarse 0.005 Fine 0.0025 | Collision distance $\alpha_1$ | 0.85 |
| Effective radius $r_e$ (m) | Small $2.1 \times l_i$ Large $3.1 \times l_i$ | Coefficient of restitution $\alpha_2$ | 0.2 |
| Time step $\Delta t$ (s) | $1.25 \times 10^{-4}$ | Surface threshold $\beta_F$ | 0.93 |
| Speed of perturbations $c_s$ (m/s) | 2.0 | Surface threshold $\varrho_F$ | 0.25 |
| Artificial compressibility $\alpha_c$ (ms²/kg) | $2.0 \times 10^{-8}$ | | |

Figure 19 displays some snapshots of the cylinder penetration and free surface profile comparing experiment and numerical simulation at the instants $t(g/D_s)^{1/2} = 0.047, 0.188,$ 0.283 and 0.793 ($t = 0.005, 0.020, 0.030$ and $0.084$s). The colors of the fluid particles are related to its non-dimensional pressure magnitude $P/\rho g D_s$. At the instant $t(g/D_s)^{1/2} = 0.047$, high pressure is computed just after the impact. After, the computed pressure near the cylinder bottom decreases and a jet flow separates from the cylinder surface at $t(g/D_s)^{1/2} = 0.188$. Subsequently, as the cylinder penetrates the fluid between $t(g/D_s)^{1/2} = 0.283$ and $0.793$, more evident splash and fragmentation of the free surface can be observed. In general, the computed evolution of the free surface is in good agreement with the experimental one. Concerning the pressure field, a remarkably smooth pattern is computed with BMMR-MPS.

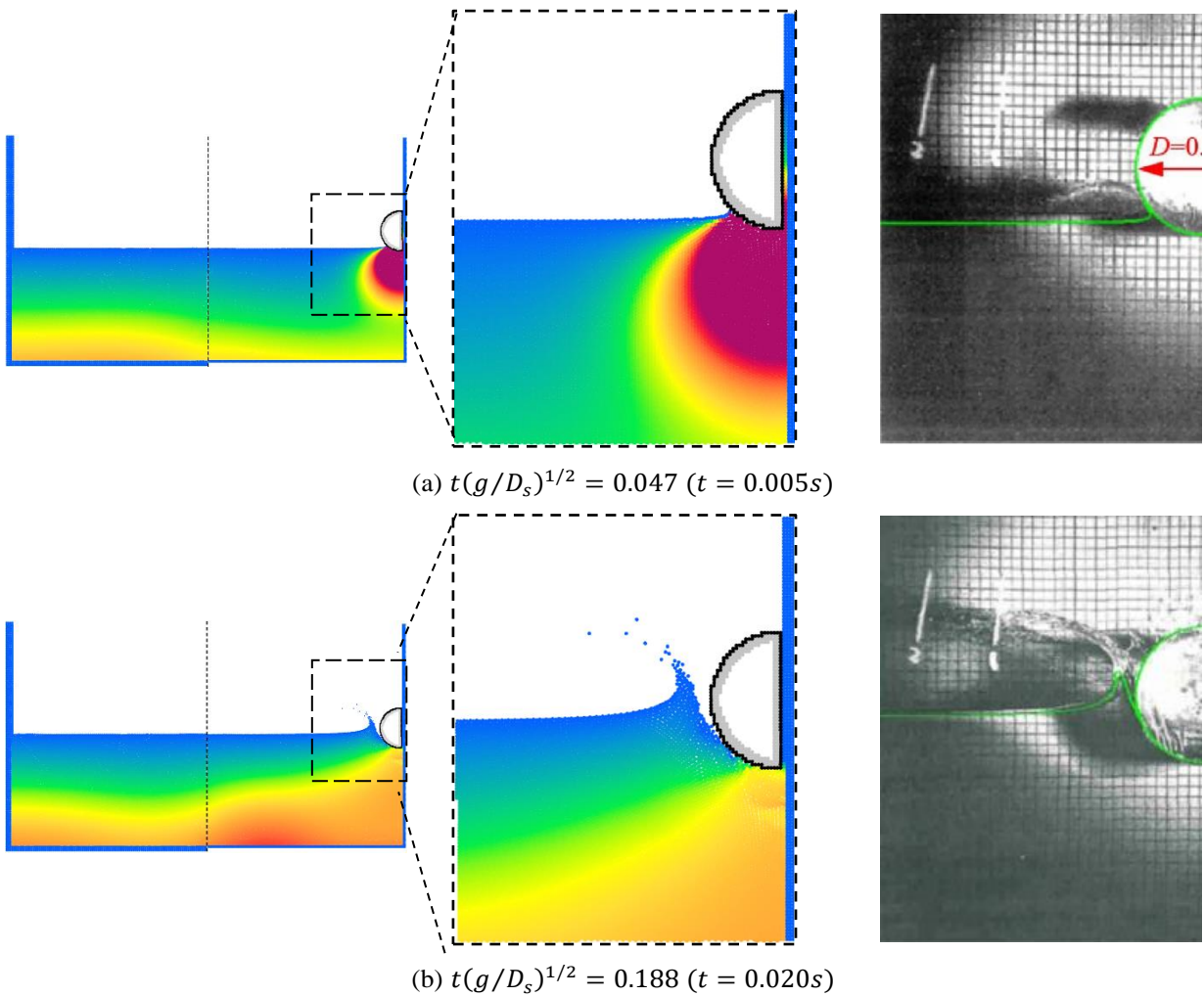

(a) $t(g/D_s)^{1/2} = 0.047$ ($t = 0.005s$)

(b) $t(g/D_s)^{1/2} = 0.188$ ($t = 0.020s$)

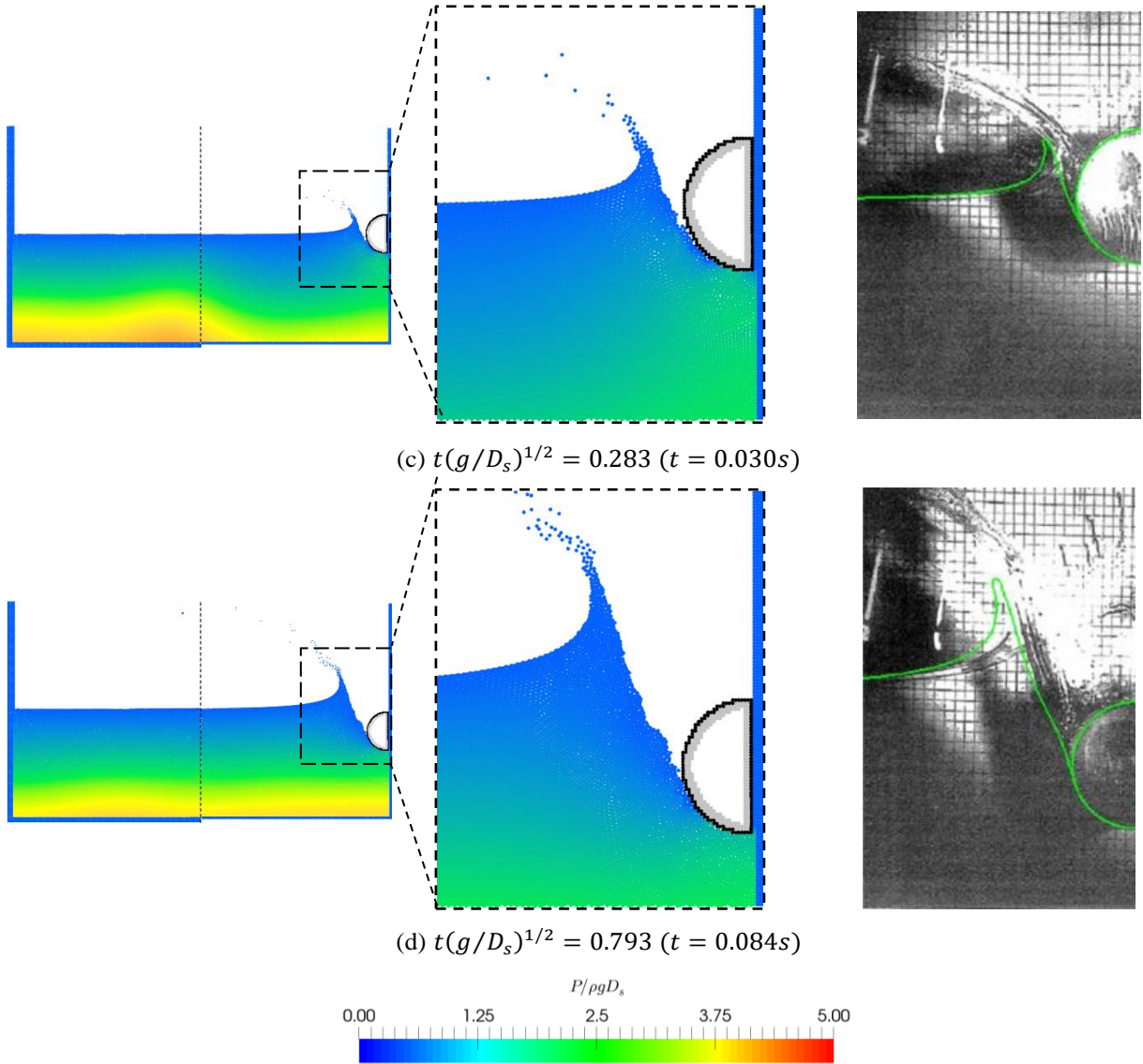

Figure 19 - Free surface profile and half buoyant circular cylinder penetration from the present multi-resolution BMMR-MPS simulation, the left symmetrical half region of the experiment (Greenhow & Lin, 1983), and (highlighted by the green solid lines) the BEM results (Sun & Faltinsen, 2006) at the instants 0.047, 0.188, 0.283, and 0.793 ($t = 0.005, 0.020, 0.030$ and $0.084s$). The colors of the fluid particles are related to its pressure.

The comparison between the depth of penetration of the circular cylinder from the experiment (Greenhow & Lin, 1983), BEM result (Sun & Faltinsen, 2006) and BMMR-MPS is shown in Figure 20. The initial time $t = 0$ is set to the moment when the cylinder contacts the water surface. The numerical results computed by BEM and BMMR-MPS agree very well. Overall, the present computed results are in good agreement with the experimental data, except for the slight overestimated experimental value near $t(g/D_s)^{1/2} = 1.18$, which can also be observed in numerical simulations conducted by (Sun, et al., 2018; Nguyen, et al., 2020; Yang, et al., 2021). The adopted numerical resolutions can be related to the difference in the experimental and computed drag forces, resulting in this slightly deviation from the experimental data.

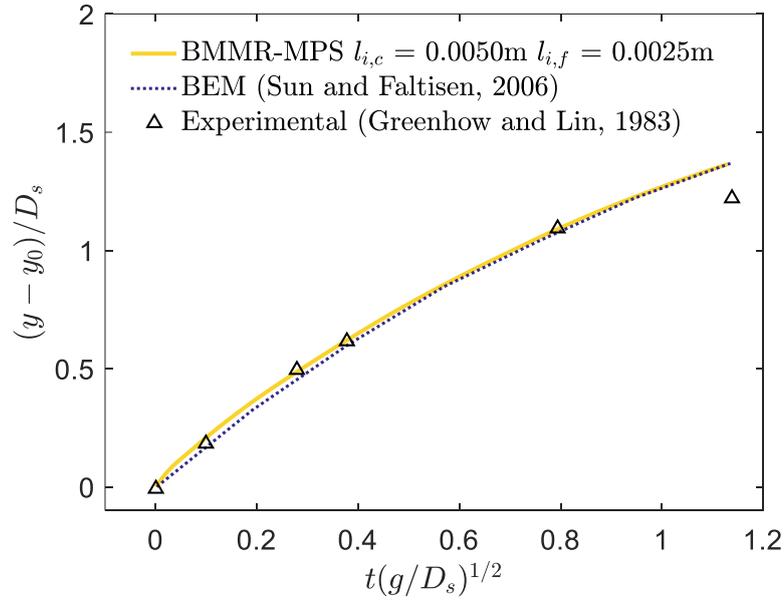

Figure 20 - Water impact of circular cylinder. Depth of penetration of the circular cylinder from the experiment (Greenhow & Lin, 1983), BEM result (Sun & Faltinsen, 2006) and the present result using the proposed multi-resolution BMMR-MPS with $l_{i,c} = 0.005$m (coarse-resolution), $l_{i,f} = 0.0025$m (fine-resolution).

## 5 Computational time solving the pressure Poisson equation (PPE)

The solution of the PPE linear system by using iterative solvers is the most time-consuming task of incompressible projection-based particle methods (Hori, et al., 2011; Fernandes, et al., 2015; Chow, et al., 2018; Guo, et al., 2018; Chen & Wan, 2019; Fourtakas, et al., 2021). To evaluate the performance of the present multi-resolution scheme in reducing the runtime to solve the PPE linear system, 2D inviscid standing wave problems, of which the main dimensions are the same used in Section 4.1, as shown in Figure 11, were simulated using different proportions of coarse and fine resolutions. The heights of coarse subdomain $H_c = 0, 0.22, 0.44,$ 0.66, 0.88 and 1.1m (height percentages of 0, 20%, 40%, 60%, 80% and 100%), and particle distances $l_i = 0.2, 0.1, 0.005$ and $0.0025$m were evaluated. Single-resolution MPS was used for $H_c = 0$, while multi-resolution was used for the remaining cases.

Since the matrix of coefficients of the single-resolution simulation is diagonal and symmetrical, the Conjugate Gradient (CG) method could be adopted. On the other hand, the matrix of coefficients of the multi-resolution simulation is not strictly diagonal neither symmetrical. Therefore, the CG solver could not be used, and a Generalized Minimal Residual (GMRES) solver was adopted instead. For both solvers, we used the maximum relative tolerance $\epsilon_{tol} = 10^{-2}$ for the stopping criterion, based on the current ($\|\mathbf{r}_k\| = \|\mathbf{Ap}_k - \mathbf{b}\|$) residual norm:

$$\frac{\|\mathbf{Ap}_k - \mathbf{b}\|}{\|\mathbf{b}\|} \leq \epsilon_{tol} . \tag{45}$$

The computational efficiency of the single-resolution and multi-resolution simulations, respectively using CG and GMRES solvers, are compared in Table 5. Simulations using the single-resolution MPS were used as reference, i.e., the speedup represents the ratio of the uniform fine single-resolution runtime to the time taken by the BMMR-MPS for the same problem. As shown in Table 5, the speedup can reach more than 2 times when 20% of the domain is discretized by fine resolution. Figure 21 shows the comparison of the computational efficiency of the solver of the standing wave simulations using uniform high resolution and multi-resolution domain. From Figure 21, the speedup decreases with the decrease of the initial particle distance. The decrease in the initial particle distance leads to an increase in number of *simplified* or *fictitious* particles close to the border of the subdomains. Since the contribution of these particles are introduced outside the main diagonal of the linear system from PPE, linear systems will become less diagonal dominant (see Figure 8). Consequently, solve the multi-resolution PPE will be more time consuming with smaller particle distance, although it still is faster than solve a single fine resolution PPE, which is diagonal and symmetric but much larger.

Table 5 - Comparison of the computational efficiency of the solver of the standing wave simulations using uniform fine resolution and multi-resolution domain.

| Particle distance (m) | | Number of particles | | | Solver | | Speedup |
|---|---|---|---|---|---|---|---|
| Coarse | Fine | Coarse | Fine | Total | Type | Runtime/step (s) | |
| 0.02 | 0.01 | 0 | 21338 | 21338 | CG | 0.0187 | 1.00 |
| | | 1500 (20%) | 16600 (80%) | 18100 | GMRES | 0.0170 | 1.10 |
| | | 2500 (40%) | 12500 (60%) | 15000 | | 0.0132 | 1.42 |
| | | 3670 (60%) | 8290 (40%) | 11960 | | 0.0100 | 1.87 |
| | | 4750 (80%) | 4110 (20%) | 8860 | | 0.0068 | 2.75 |
| 0.01 | 0.005 | 0 | 82660 | 82660 | CG | 0.063 | 1.00 |
| | | 5000 (20%) | 65300 (80%) | 70300 | GMRES | 0.059 | 1.07 |
| | | 9150 (40%) | 48950 (60%) | 58100 | | 0.050 | 1.26 |
| | | 13300 (60%) | 32600 (40%) | 45900 | | 0.036 | 1.75 |
| | | 17470 (80%) | 16230 (20%) | 33700 | | 0.025 | 2.52 |
| 0.005 | 0.0025 | 0 | 325300 | 325300 | CG | 0.26 | 1.00 |
| | | 18350 (20%) | 258650 (80%) | 277000 | GMRES | 0.26 | 1.00 |
| | | 34670 (40%) | 193905 (60%) | 228575 | | 0.21 | 1.24 |
| | | 51000 (60%) | 129175 (40%) | 180175 | | 0.19 | 1.37 |
| | | 67310 (80%) | 64465 (20%) | 131775 | | 0.13 | 2.00 |

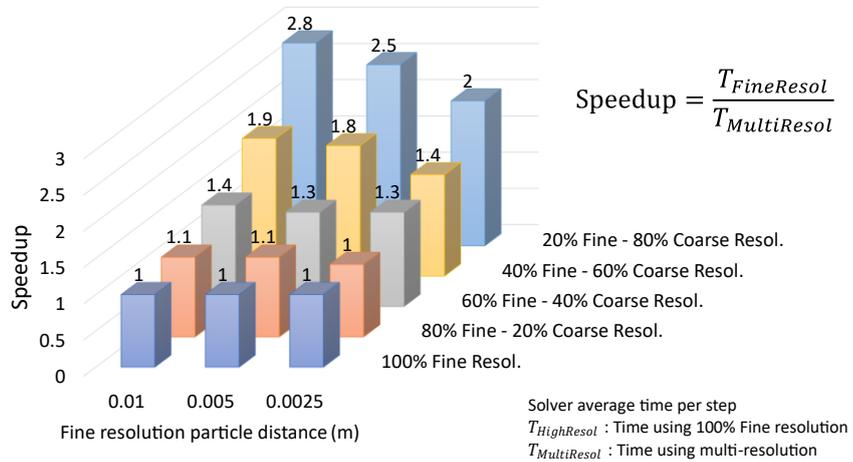

Figure 21 - Comparison of the computational efficiency of the solver of the standing wave simulations using uniform fine resolution and multi-resolution domain.

# 6 Concluding remarks

A multi-resolution technique named border mapping multi-resolution (BMMR) was proposed for projection-based particle methods. In the BMMR, a background equivalent particle distribution is obtained in the truncated border between subdomains with different resolutions by means of two procedures: the refinement of the coarse resolution particle distribution and the simplification of the fine resolution particle distribution. Moreover, the technique provides a strong coupling between the subdomains of different resolutions. To solve the governing equations of incompressible free surface flows, the moving particle semi-implicit (MPS) method was adopted, resulting in the proposed border mapping multi-resolution MPS (BMMR-MPS).

By conducting simulations of 2D standing wave, the conservation of the global properties such as mass and mechanical energy was verified. The simulations of 2D dam break and water impact of circular cylinder show that the proposed approach is numerically stable and can predict the flow evolution, pressure field, and fluid-solid interaction very well.

The performance of the BMMR-MPS in reducing the runtime to solve the PPE linear system was assessed by simulating 2D inviscid standing wave problems by using different proportions of the computational domain as coarse and fine resolutions. It was shown that the proposed model can reduce the computational costs significantly, reaching speedups of around 2 times compared to uniform single-resolution simulations.

As next steps and future works, since the BMMR technique is suitable for any projection-based particle method, it could be adopted for the incompressible smoothed particle hydrodynamics (ISPH) or consistent particle method (CPM) (Koh, et al., 2012) as well. On the other hand, the border refinement and simplification algorithms were executed sequentially in this first implementation, leading to a high computational cost. In this sense, we will further improve the performance of the proposed method, taking advantage of the parallelization by, e.g., Open

Multi-Processing (OpenMP)[3] or Intel® Threading Building Blocks (TBB)[4] directives. More efficient refinement and simplification techniques might also be investigated. Moreover, as further extension of BMMR to fully exploit the potential of the technique, multi-level refinement, subdomains delimited by curved borders and adaptive border might also be considered. Finally, the border-mapping technique should be developed for the 3D formulation, which will be substantially more challenging regarding the refinement and simplification algorithms.

**CRediT authorship contribution statement**

**Cezar Augusto Bellezi**: Conceptualization of this study, Methodology, Software, Formal analysis, Writing – original draft. **Liang-Yee Cheng**: Conceptualization of this study, Methodology, Writing – review & editing, Supervision, Funding acquisition. **Rubens Augusto Amaro Jr**: Methodology, Software, Validation, Formal analysis, Writing – original draft. **Marcio Michiharu Tsukamoto**: Software, Formal analysis.

**Acknowledgements**

This material is based upon research supported by the Office of Naval Research Global under the Award Number: N62909-16-1-2181. The first and third authors would like to thank the Coordenação de Aperfeiçoamento de Pessoal de Nível Superior - Brasil (CAPES) - Finance Code 001, for the doctorate scholarships. The authors are also grateful to Petrobras for financial support on the development of the MPS/TPN-USP simulation system.

**References**

Altomare, C. et al., 2018. Improved relaxation zone method in SPH-based model for coastal engineering applications. *Applied Ocean Research,* Volume 81, pp. 15-33.

Antuono, M., Colagrossi, A., Marrone, S. & Lugni, C., 2011. Propagation of gravity waves

---

[3] http://www.openmp.org
[4] https://software.intel.com/content/www/us/en/develop/tools/oneapi/components/onetbb.html

through an SPH scheme with numerical diffusive terms. *Computer Physics Communications,* 182(4), pp. 866-877.

Arai, J., Koshizuka, S. & Murozono, K., 2013. Large eddy simulation and a simple wall model for turbulent flow calculation by a particle method. *International Journal for Numerical Methods in Fluids,* 71(6), pp. 772-787.

Barcarolo, D. A., Le Touzé, D., Oger, G. & De Vuyst, F., 2014. Adaptive particle refinement and derefinement applied to the smoothed particle hydrodynamics method. *Journal of Computational Physics,* Volume 273, pp. 640-657.

Cheng, L.-Y., Amaro Junior, R. A. & Favero, E. H., 2021. Improving stability of moving particle semi-implicit method by source terms based on time-scale correction of particle-level impulses. *Engineering Analysis with Boundary Elements,* Volume 131, pp. 118-145.

Chen, X., Sun, Z. G., Liu, L. & Xi, G., 2016. Improved MPS method with variable-size particles. *International Journal for Numerical Methods in Fluids,* 80(6), pp. 358-374.

Chen, X. & Wan, D., 2019. GPU accelerated MPS method for large-scale 3-D violent free surface flows. *Ocean Engineering,* Volume 171, pp. 677-694.

Chen, X., Xi, G. & Sun, Z.-G., 2014. Improving stability of MPS method by a computational scheme based on conceptual particles. *Computer Methods in Applied Mechanics and Engineering*, Volume 278, pp. 254-271.

Chiron, L., Marrone, S., Di Mascio, A. & Le Touzé, D., 2018. Coupled SPH–FV method with net vorticity and mass transfer. *Journal of Computational Physics,* Volume 364, pp. 111-136.

Chiron, L., Oger, G., de Leffeb, M. & Le Touzé, D., 2018. Analysis and improvements of Adaptive Particle Refinement (APR) through CPU time, accuracy and robustness considerations. *Journal of Computational Physics,* Volume 354, pp. 552-575.

Chorin, A. J., 1967. The numerical solution of the Navier-Stokes equations for an incompressible fluid. *Bulletin of the American Mathematical Society,* 73(6), pp. 928-931.

Chow, A. D., Rogers, B. D., Lind, S. J. & Stansby, P. K., 2018. Incompressible SPH (ISPH)

with fast Poisson solver on a GPU. *Computer Physics Communications,* Volume 226, pp. 81-103.

Courant, R., Friedrichs, K. & Levy, H., 1967. On the partial difference equations of mathematical physics. *IBM Journal of Research and Development,* 11(2), pp. 215-234.

Cummins, S. J. & Rudman, M., 1999. An SPH projection method. *Journal of Computational Physics,* 152(2), pp. 584-607.

Di Mascio, A. et al., 2021. SPH-FV coupling algorithm for solving multi-scale three-dimensional free-surface flows. *Applied Ocean Research,* Volume 115.

Duan, G., Yamaji, A., Koshizuka, S. & Chen, B., 2019. The truncation and stabilization error in multiphase moving particle semi-implicit method based on corrective matrix: Which is dominant ?. *Computers & Fluids,* Volume 190, pp. 254-273.

Feldman, J. & Bonet, J., 2007. Dynamic refinement and boundary contact forces in SPH with applications in fluid flow problems. *International Journal for Numerical Methods in Engineering,* 72(3), pp. 295-324.

Fernandes, D. T., Cheng, L.-Y., Favero, E. H. & Nishimoto, K., 2015. A domain decomposition strategy for hybrid parallelization of moving particle semi-implicit (MPS) method for computer cluster. *Cluster Computing,* 18(4), pp. 1363-1377.

Fourtakas, G., Rogers, B. D. & Nasar, A. M. A., 2021. Towards pseudo-spectral incompressible smoothed particle hydrodynamics (ISPH). *Computer Physics Communications,* Volume 266.

Gingold, R. A. & Monaghan, J. J., 1977. Smoothed particle hydrodynamics: theory and application to non-spherical stars. *Monthly Notices of the Royal Astronomical Society*, Volume 181, pp. 375-389.

Gotoh, H. & Khayyer, A., 2018. On the state-of-the-art of particle methods for coastal and ocean engineering. *Coastal Engineering Journal,* 60(1), pp. 79-103.

Gotoh, H., Okayasu, A. & Watanabe, Y., 2013. *Computational wave dynamics.* s.l.:World Scientific Publishing Company.

Greenhow, M. & Lin, W.-M., 1983. *Nonlinear free surface effects: experiments and theory.* Cambridge: Massachusetts Inst Tech.

Guo, X., Rogers, B. D., Lind, S. & Stansby, P. K., 2018. New massively parallel scheme for incompressible smoothed particle hydrodynamics (ISPH) for highly nonlinear and distorted flow. *Computer Physics Communications,* Volume 233, pp. 16-28.

Harlow, F. H. & Welch, J. E., 1965. Numerical calculation of time-dependent viscous incompressible flow of fluid with a free surface. *The Physics of Fluids,* 8(12), pp. 2182-2189.

Henshaw, W. D. & Kreiss, H.-O., 1995. *Analysis of a difference approximation for the incompressible Navier-Stokes equations,* s.l.: s.n.

Hori, C., Gotoh, H., Ikari, H. & Khayyer, A., 2011. GPU-acceleration for moving particle semi-Implicit method. *Computers & Fluids,* 51(1), pp. 174-183.

Hu, L., HongFu, Q., FuZhen, C. & Chao, S., 2020. A particle refinement scheme with hybrid particle interacting technique for multi-resolution SPH. *Engineering Analysis with Boundary Elements,* Volume 118, pp. 108-123.

Jandaghian, M. & Shakibaeinia, A., 2020. An enhanced weakly-compressible MPS method for free-surface flows. *Computer Methods in Applied Mechanics and Engineering,* Volume 360.

Khayyer, A. & Gotoh, H., 2013. Enhancement of performance and stability of MPS mesh-free particle method for multiphase flows characterized by high density ratios. *Journal of Computational Physics,* Volume 242, pp. 211-233.

Khayyer, A., Gotoh, H., Shimizu, Y. & Gotoh, K., 2017. On enhancement of energy conservation properties of projection-based particle methods. *European Journal of Mechanics - B/Fluids,* Volume 66, pp. 20-37.

Khayyer, A., Tsuruta, N., Shimizu, Y. & Gotoh, H., 2019. Multi-resolution MPS for incompressible fluid-elastic structure interactions in ocean engineering. *Applied Ocean Research,* Volume 82, pp. 397-414.

Koh, C. G., Gao, M. & Luo, C., 2012. A new particle method for simulation of incompressible free surface flow problems. *International Journal for Numerical Methods in Engineering*, 89(12), pp. 1582-1604.

Kondo, M. & Koshizuka, S., 2011. Improvement of stability in moving particle semi-implicit method. *International Journal for Numerical Methods in Fluids,* 65(6), pp. 638-654.

Koshizuka, S. & Oka, Y., 1996. Moving-particle semi-implicit method for fragmentation of incompressible fluid. *Nuclear Science and Engineering,* 123(3), pp. 421-434.

Koshizuka, S., Shibata, K., Kondo, M. & Matsunaga, T., 2018. *Moving Particle Semi-implicit Method: A Meshfree Particle Method for Fluid Dynamics.* s.l.:Academic Press.

Lee, B.-H., Park, J.-C., Kim, M.-H. & Hwang, S.-C., 2011. Step-by-step improvement of MPS method in simulating violent free-surface motions and impact-loads. *Computer Methods in Applied Mechanics and Engineering,* 200(9-12), pp. 1113-1125.

Lee, E.-S.et al., 2008. Comparisons of weakly compressible and truly incompressible algorithms for the SPH mesh free particle method. *Journal of Computational Physics,* 227(18), pp. 8417-8436.

Li, G. et al., 2020. A review on MPS method developments and applications in nuclear engineering. *Computer Methods in Applied Mechanics and Engineering,* Volume 367.

Li, L., 2020. A split-step finite-element method for incompressible Navier-Stokes equations with high-order accuracy up-to the boundary. *Journal of Computational Physics,* Volume 408.

Lind, S. J., Rogers, B. D. & Stansby, P. K., 2020. Review of smoothed particle hydrodynamics: towards converged Lagrangian flow modelling. *Proceedings of the Royal Society A,* 476(2241).

Lind, S. J., Xu, R., Stansby, P. K. & Rogers, B. D., 2012. Incompressible smoothed particle hydrodynamics for free-surface flows: a generalised diffusion-based algorithm for stability and validations for impulsive flows and propagating waves. *Journal of Computational Physics*, 231(4), pp. 1499-1523.

Liu, X. & Zhang, S., 2021. Development of adaptive multi-resolution MPS method for multiphase flow simulation. *Computer Methods in Applied Mechanics and Engineering,*

Volume 387.

Lobovský, L. et al., 2014. Experimental investigation of dynamic pressure loads during dam break. *Journal of Fluids and Structures,* Volume 48, pp. 407-434.

Lo, E. Y. M. & Shao, S., 2002. Simulation of near-shore solitary wave mechanics by an incompressible SPH method. *Applied Ocean Research,* 24(5), pp. 275-286.

Lucy, L. B., 1977. A numerical approach to the testing of the fission hypothesis. *Astronomical Journal,* Volume 82, pp. 1013-1024.

Luo, M., Khayyer, A. & Lin, P., 2021. Particle methods in ocean and coastal engineering. *Applied Ocean Research,* Volume 114.

Lyu, H.-G., Deng, R., Sun, P.-N. & Miao, J.-M., 2021. Study on the wedge penetrating fluid interfaces characterized by different density-ratios: Numerical investigations with a multi-phase SPH model. *Ocean Engineering,* Volume 237.

Matsunaga, T., Södersten, A., Shibata, K. & Koshizuka, S., 2020. Improved treatment of wall boundary conditions for a particle method with consistent spatial discretization. *Computer Methods in Applied Mechanics and Engineering,* Volume 358.

Monaghan, J. J., 1994. Simulating free surface flows with SPH. *Journal of Computational Physics*, Volume 110, pp. 399-406.

Nguyen, V. T., Phan, T. H. & Park, W. G., 2020. Modeling and numerical simulation of ricochet and penetration of water entry bodies using an efficient free surface model. *International Journal of Mechanical Sciences,* p. 182.

Reyes López, Y., Roose, D. & Recarey Morfa, C., 2013. Dynamic particle refinement in SPH: application to free surface flow and non-cohesive soil simulations. *Computational Mechanics*, Volume 51, pp. 731-741.

Riley, J. D., 1955. Solving systems of linear equations with a positive definite, symmetric, but possibly ill-conditioned matrix. *Mathematical Tables and Other Aids to Computation,* 9(51), pp. 96-101.

Shibata, K., Koshizuka, S., Matsunaga, T. & Massaie, I., 2017. The overlapping particle technique for multi-resolution simulation of particle methods. *Computer Methods in Applied Mechanics and Engineering,* Volume 325, pp. 434-462.

Shibata, K., Koshizuka, S., Tamai, T. & Murozono, K., 2012. *Overlapping particle technique and application to green water on deck.* Nantes, France, s.n.

Shibata, K., Masaie, I., Kondo, M. & Murotani, K., 2015. Improved pressure calculation for the moving particle semi-implicit method. *Computational Particle Mechanics,* Volume 2, pp. 91-108.

Shobeyri, G. & Ardakani, R. R., 2019. Modified incompressible SPH method for simulating free surface problems using highly irregular multi-resolution particle configurations. *Journal of the Brazilian Society of Mechanical Sciences and Engineering,* 41(10), pp. 1-15.

Skillen, A., Lind, S., Stansby, P. K. & Rogers, B. D., 2013. Incompressible smoothed particle hydrodynamics (SPH) with reduced temporal noise and generalised Fickian smoothing applied to body–water slam and efficient wave–body interaction. *Computer Methods in Applied Mechanics and Engineering,* Volume 265, pp. 163-173.

Souto-Iglesias, A., Macià, F., González, L. M. & Cercos-Pita, J. C., 2013. On the consistency of MPS. *Computer Physics Communications,* 184(3), pp. 732-745.

Sun, H. & Faltinsen, O. M., 2006. Water impact of horizontal circular cylinders and cylindrical shells. *Applied Ocean Research,* 28(5), pp. 299-311.

Sun, P., Zhang, A. M., Marrone, S. & Ming, F., 2018. An accurate and efficient SPH modeling of the water entry of circular cylinders. *Applied Ocean Research,* Volume 72, pp. 60-75.

Sun, Z., Djidjeli, K. & Xing, J. T., 2017. The weak coupling between MPS and BEM for wave structure interaction simulation. *Engineering Analysis with Boundary Elements,* Volume 82, pp. 111-118.

Suzuki, Y., Koshizuka, S. & Oka, Y., 2007. Hamiltonian moving-particle semi-implicit (HMPS) method for incompressible fluid flows. *Computer Methods in Applied Mechanics and Engineering,* 196(29-30), pp. 2876-2894.

Tamai, T. & Koshizuka, S., 2014. Least squares moving particle semi-implicit method. *Computational Particle Mechanics*, September, 1(3), pp. 277-305.

Tanaka, M., Cardoso, R. & Bahai, H., 2018. Multi-resolution MPS method. *Journal of Computational Physics,* Volume 359, pp. 106-136.

Tanaka, M. & Masunaga, T., 2010. Stabilization and smoothing of pressure in MPS method by Quasi-Compressibility. *Journal of Computational Physics,* 229(11), pp. 4279-4290.

Tanaka, M., Masunaga, T. & Nakagawa, Y., 2009. Multi-resolution MPS Method. *Transactions of the Japanese Socieety of Computational Engineering and Science (JSCES) (in japanese)*.

Tang, Z., Wan, D., Chen, G. & Xiao, Q., 2016b. Numerical simulation of 3D violent free-surface flows by multi-resolution MPS method. *Journal of Ocean Engineering and Marine Energy,* Volume 2, pp. 355-364.

Tang, Z., Zhang, Y. & Wan, D., 2016a. Multi-Resolution MPS Method for Free Surface Flows. *International Journal of Computational Methods,* 13(4).

Tang, Z., Zhang, Y. & Wan, D., 2016c. Numerical simulation of 3D free surface flows by overlapping MPS. *Journal of Hydrodynamics*, Volume 28 (2), pp. 306-312.

Temam, R., 1969. Sur l'approximation de la solution des équations de Navier-Stokes par la méthode des pas fractionnaires (II). *Archive for Rational Mechanics and Analysis,* 33(5), pp. 377-385.

Tsukamoto, M. M. et al., 2020. A numerical study of the effects of bottom and sidewall stiffeners on sloshing behavior considering roll resonant motion. *Marine Structures,* Volume 72.

Tsukamoto, M. M., Cheng, L.-Y. & Motezuki, F. K., 2016. Fluid interface detection technique based on neighborhood particles centroid deviation (NPCD) for particle methods. *International Journal for Numerical Methods in Fluids,* 82(3), pp. 148-168.

Tsuruta, N., Gotoh, H. & Khayyer, A., 2016. *A novel refinement technique for projection-based particle methods.* Munich, Germany, s.n., pp. 402-410.

Tsuruta, N., Khayyer, A. & Gotoh, H., 2015. Space potential particles to enhance the stability of projection-based particle methods. *International Journal of Computational Fluid Dynamics,* 29(1), pp. 100-119.

Vacondio, R. et al., 2013. Variable resolution SPH: a dynamic particle coalescing and splitting scheme. *Computer Methods in Applied Mechanics and Engineering,* Volume 256, pp. 132-148.

Vacondio, R., Rogers, B. D., Stansby, P. K. & Mignosa, P., 2016. Variable resolution SPH in three dimensions: towards optimal splitting and coalescing for dynamic adaptivity. *Computer Methods in Applied Mechanics and Engineering*, Volume 300, pp. 442-460.

Verbrugghe, T. et al., 2018. Coupling methodology for smoothed particle hydrodynamics modelling of non-linear wave-structure interactions. *Coastal Engineering,* Volume 138, pp. 184-198.

Wang, L., Jiang, Q. & Zhang, C., 2017. Improvement of moving particle semi-implicit method for simulation of progressive water waves. *International Journal for Numerical Methods in Fluids,* 85(2), pp. 69-89.

Wu, G. X. & Taylor, R. E., 1994. Finite element analysis of two-dimensional non-linear transient water waves. *Applied Ocean Research,* 16(6), pp. 363-372.

Yang, X., Kong, S.-C., Liu, M. & Liu, Q., 2021. Smoothed particle hydrodynamics with adaptive spatial resolution (SPH-ASR) for free surface flows. *Journal of Computational Physics*.

Ye, T., Pan, D., Huang, C. & Liu, M., 2019. Smoothed particle hydrodynamics (SPH) for complex fluid flows: Recent developments in methodology and applications. *Physics of Fluids,* 31(1).